\documentstyle[11pt]{article}

\makeatletter

\ifcase \@ptsize
\or 
\or 
\fi
\advance\textheight by \topskip
\ifcase \@ptsize
\or 
\or 
\fi
\def\icite{\@ifnextchar [{\@tempswatrue\@citey}{\@tempswafalse\@citey[]}}
\def\@citex[#1]#2{%
\if@filesw \immediate \write \@auxout {\string \citation {#2}}\fi
\@tempcntb\m@ne \let\@h@ld\relax \def\@citea{}%
\@cite{%
  \@for \@citeb:=#2\do {%
    \@ifundefined {b@\@citeb}%
      {\@h@ld\@citea\@tempcntb\m@ne{\bf ?}%
      \@warning {Citation `\@citeb ' on page \thepage \space undefined}}%
      {\@tempcnta\@tempcntb \advance\@tempcnta\@ne%
      \@tempcntb\number\csname b@\@citeb \endcsname \relax%
      \ifnum\@tempcnta=\@tempcntb 
        \ifx\@h@ld\relax%
          \edef \@h@ld{\@citea\csname b@\@citeb\endcsname}%
        \else%
          \edef\@h@ld{\ifmmode{-}\else--\fi\csname b@\@citeb\endcsname}%
        \fi%
      \else
        \@h@ld\@citea\csname b@\@citeb \endcsname%
        \let\@h@ld\relax%
      \fi}%
    \def\@citea{,\penalty\@highpenalty\,}%
  }\@h@ld
}{#1}}
\def\@citey[#1]#2{%
\if@filesw \immediate \write \@auxout {\string \citation {#2}}\fi
\@tempcntb\m@ne \let\@h@ld\relax \def\@citea{}%
\@icite{%
  \@for \@citeb:=#2\do {%
    \@ifundefined {b@\@citeb}%
      {\@h@ld\@citea\@tempcntb\m@ne{\bf ?}%
      \@warning {Citation `\@citeb ' on page \thepage \space undefined}}%
      {\@tempcnta\@tempcntb \advance\@tempcnta\@ne%
      \@tempcntb\number\csname b@\@citeb \endcsname \relax%
      \ifnum\@tempcnta=\@tempcntb 
        \ifx\@h@ld\relax%
          \edef \@h@ld{\@citea\csname b@\@citeb\endcsname}%
        \else%
          \edef\@h@ld{\ifmmode{-}\else--\fi\csname b@\@citeb\endcsname}%
        \fi%
      \else
        \@h@ld\@citea\csname b@\@citeb \endcsname%
        \let\@h@ld\relax%
      \fi}%
    \def\@citea{,\penalty\@highpenalty\,}%
  }\@h@ld
}{#1}}
\def\@cite#1#2{{$\!\! ^{#1}$\if@tempswa , #2\fi }}
\def\@icite#1#2{{#1\if@tempswa , #2\fi }}
\def\thebibliography#1{\section*{References\@mkboth
 {REFERENCES}{REFERENCES}}\list
 {\arabic{enumi}.}{\settowidth\labelwidth{#1.}\leftmargin\labelwidth
 \advance\leftmargin\labelsep
 \usecounter{enumi}}
 \def\newblock{\hskip .11em plus .33em minus .07em}
 \sloppy\clubpenalty4000\widowpenalty4000
 \sfcode`\.=1000\relax}
\font\twelveitb = cmbxti10 scaled \magstep1%

\@addtoreset{equation}{section}
\def\section{\@startsection {section}{1}{\z@}{-3.5ex plus -1ex minus
 -.2ex}{2.3ex plus .2ex}{\normalsize\bf}}
\def\subsection{\@startsection{subsection}{2}{\z@}{-3.25ex plus -1ex minus
 -.2ex}{1.5ex plus .2ex}{\normalsize\twelveitb}}
\def\subsubsection{\@startsection{subsubsection}{3}{\z@}{-3.25ex plus
 -1ex minus -.2ex}{1.5ex plus .2ex}{\normalsize\rm}}
\def\acknowledgements{\@startsection{section}{4}
{\z@}{-3.5ex plus -1ex minus -.2ex}{2.3ex plus .2ex}{\normalsize\bf}
{Acknowledgements}}

\def\thefootnote {\alph{footnote}}
\gdef\@publabel{\hfil}
\gdef\@pubdate{\null}
\gdef\@pubnumber{\null}
\gdef\@author{\null}
\gdef\@title{\null}
\gdef\@abstract{\null}
\long\def\pubdate#1{\gdef\@pubdate{#1}}
\long\def\pubnumber#1{\gdef\@pubnumber{#1}}
\long\def\publabel#1{\gdef\@publabel{#1}}
\long\def\author#1{\gdef\@author{#1}}
\long\def\title#1{\gdef\@title{#1}}
\long\def\abstract#1{\gdef\@abstract{#1}}
\def\titlerelax{
\let\maketitle\relax
\let\settitleparameters\relax
\let\consolidatetitle\relax
\let\inittitlepage\relax
\let\finishtitlepage\relax
\let\titlepagecontents\relax
\let\multithanks\relax
\let\titlebaselines\relax
\let\@makepub\relax
\let\@maketitle\relax
\let\@makeauthor\relax
\let\@makeabstract\relax
\let\@maketitlenote\relax
\let\thanks\relax
\let\titlerelax\relax}
\def\titleclean
{\gdef\@titlenote{}
\gdef\@abstract{}
\gdef\@author{}
\gdef\@title{}
\gdef\@pubdate{}\gdef\@pubnumber{}\gdef\@publabel{}
\gdef\@dpublabel{}
}

\def\@makepub{\vbox to \z@{\hbox to \textwidth{\hfill
\@publabel \hfill
\llap{\parbox[t]{0.25\textwidth}{\raggedleft\@pubnumber}}}%
\vss}}
\def\@maketitle{\vskip 60pt \begin{center}
 {\LARGE \@title \par}
 \end{center}}
\def\@makeauthor{{%
\def\and{\smallskip {\normalsize \rm and\smallskip }}
\def\And{\medskip {\normalsize \rm and\\}\medskip}
\long\def\address##1{{\def\and{\\and\\}\medskip
				{\small \it \\##1\\}
}}
{\centering
 \vskip 3em
 \large \lineskip .75em
 \@author}
 \par}}
\def\@makedate{\vskip 1.5em
 {\raggedright \small \noindent\@pubdate \par}}
\def\@makeabstract{\vskip 1.5em
{\small
\begin{center}
{\bf ABSTRACT\vspace{-.5em}\vspace{0pt}}
\end{center}
\quotation \@abstract \endquotation}}
\def\maketitle{\titlepage
\let\footnotesize\small \setcounter{page}{0}
\def\thefootnote{\fnsymbol{footnote}}
\@makepub
\vfil
\@maketitle
\@makeauthor
\vfil
\@makeabstract
\@thanks
\vfil
\@makedate
\if@restonecol\twocolumn \else \eject \fi
\titlerelax \titleclean
\def\thefootnote{\alph{footnote}}
\setcounter{footnote}{0}
}

\ifcase\@ptsize
 \font\tenmsa=msam10
 \font\sevenmsa=msam7
 \font\fivemsa=msam5
 \font\tenmsb=msbm10
 \font\sevenmsb=msbm7
 \font\fivemsb=msbm5
\or
 \font\tenmsa=msam10 scaled \magstephalf
 \font\sevenmsa=msam8
 \font\fivemsa=msam6
 \font\tenmsb=msbm10 scaled \magstephalf
 \font\sevenmsb=msbm8
 \font\fivemsb=msbm6
\or
 \font\tenmsa=msam10 scaled \magstep1
 \font\sevenmsa=msam8
 \font\fivemsa=msam6
 \font\tenmsb=msbm10 scaled \magstep1
 \font\sevenmsb=msbm8
 \font\fivemsb=msbm6
\fi

\newfam\msafam
\newfam\msbfam
\textfont\msafam=\tenmsa  \scriptfont\msafam=\sevenmsa
  \scriptscriptfont\msafam=\fivemsa
\textfont\msbfam=\tenmsb  \scriptfont\msbfam=\sevenmsb
  \scriptscriptfont\msbfam=\fivemsb

\def\hexnumber@#1{\ifnum#1<10 \number#1\else
 \ifnum#1=10 A\else\ifnum#1=11 B\else\ifnum#1=12 C\else
 \ifnum#1=13 D\else\ifnum#1=14 E\else\ifnum#1=15 F\fi\fi\fi\fi\fi\fi\fi}

\def\msa@{\hexnumber@\msafam}
\def\msb@{\hexnumber@\msbfam}
\mathchardef\boxdot="2\msa@00
\mathchardef\boxplus="2\msa@01
\mathchardef\boxtimes="2\msa@02
\mathchardef\square="0\msa@03
\mathchardef\blacksquare="0\msa@04
\mathchardef\centerdot="2\msa@05
\mathchardef\lozenge="0\msa@06
\mathchardef\blacklozenge="0\msa@07
\mathchardef\circlearrowright="3\msa@08
\mathchardef\circlearrowleft="3\msa@09
\mathchardef\rightleftharpoons="3\msa@0A
\mathchardef\leftrightharpoons="3\msa@0B
\mathchardef\boxminus="2\msa@0C
\mathchardef\Vdash="3\msa@0D
\mathchardef\Vvdash="3\msa@0E
\mathchardef\vDash="3\msa@0F
\mathchardef\twoheadrightarrow="3\msa@10
\mathchardef\twoheadleftarrow="3\msa@11
\mathchardef\leftleftarrows="3\msa@12
\mathchardef\rightrightarrows="3\msa@13
\mathchardef\upuparrows="3\msa@14
\mathchardef\downdownarrows="3\msa@15
\mathchardef\upharpoonright="3\msa@16

\mathchardef\downharpoonright="3\msa@17
\mathchardef\upharpoonleft="3\msa@18
\mathchardef\downharpoonleft="3\msa@19
\mathchardef\rightarrowtail="3\msa@1A
\mathchardef\leftarrowtail="3\msa@1B
\mathchardef\leftrightarrows="3\msa@1C
\mathchardef\rightleftarrows="3\msa@1D
\mathchardef\Lsh="3\msa@1E
\mathchardef\Rsh="3\msa@1F
\mathchardef\rightsquigarrow="3\msa@20
\mathchardef\leftrightsquigarrow="3\msa@21
\mathchardef\looparrowleft="3\msa@22
\mathchardef\looparrowright="3\msa@23
\mathchardef\circeq="3\msa@24
\mathchardef\succsim="3\msa@25
\mathchardef\gtrsim="3\msa@26
\mathchardef\gtrapprox="3\msa@27
\mathchardef\multimap="3\msa@28
\mathchardef\therefore="3\msa@29
\mathchardef\because="3\msa@2A
\mathchardef\doteqdot="3\msa@2B

\mathchardef\triangleq="3\msa@2C
\mathchardef\precsim="3\msa@2D
\mathchardef\lesssim="3\msa@2E
\mathchardef\lessapprox="3\msa@2F
\mathchardef\eqslantless="3\msa@30
\mathchardef\eqslantgtr="3\msa@31
\mathchardef\curlyeqprec="3\msa@32
\mathchardef\curlyeqsucc="3\msa@33
\mathchardef\preccurlyeq="3\msa@34
\mathchardef\leqq="3\msa@35
\mathchardef\leqslant="3\msa@36
\mathchardef\lessgtr="3\msa@37
\mathchardef\backprime="0\msa@38
\mathchardef\risingdotseq="3\msa@3A
\mathchardef\fallingdotseq="3\msa@3B
\mathchardef\succcurlyeq="3\msa@3C
\mathchardef\geqq="3\msa@3D
\mathchardef\geqslant="3\msa@3E
\mathchardef\gtrless="3\msa@3F
\mathchardef\sqsubset="3\msa@40
\mathchardef\sqsupset="3\msa@41
\mathchardef\vartriangleright="3\msa@42
\mathchardef\vartriangleleft="3\msa@43
\mathchardef\trianglerighteq="3\msa@44
\mathchardef\trianglelefteq="3\msa@45
\mathchardef\bigstar="0\msa@46
\mathchardef\between="3\msa@47
\mathchardef\blacktriangledown="0\msa@48
\mathchardef\blacktriangleright="3\msa@49
\mathchardef\blacktriangleleft="3\msa@4A
\mathchardef\vartriangle="3\msa@4D
\mathchardef\blacktriangle="0\msa@4E
\mathchardef\triangledown="0\msa@4F
\mathchardef\eqcirc="3\msa@50
\mathchardef\lesseqgtr="3\msa@51
\mathchardef\gtreqless="3\msa@52
\mathchardef\lesseqqgtr="3\msa@53
\mathchardef\gtreqqless="3\msa@54
\mathchardef\Rrightarrow="3\msa@56
\mathchardef\Lleftarrow="3\msa@57
\mathchardef\veebar="2\msa@59
\mathchardef\barwedge="2\msa@5A
\mathchardef\doublebarwedge="2\msa@5B
\mathchardef\angle="0\msa@5C
\mathchardef\measuredangle="0\msa@5D
\mathchardef\sphericalangle="0\msa@5E
\mathchardef\varpropto="3\msa@5F
\mathchardef\smallsmile="3\msa@60
\mathchardef\smallfrown="3\msa@61
\mathchardef\Subset="3\msa@62
\mathchardef\Supset="3\msa@63
\mathchardef\Cup="2\msa@64

\mathchardef\Cap="2\msa@65

\mathchardef\curlywedge="2\msa@66
\mathchardef\curlyvee="2\msa@67
\mathchardef\leftthreetimes="2\msa@68
\mathchardef\rightthreetimes="2\msa@69
\mathchardef\subseteqq="3\msa@6A
\mathchardef\supseteqq="3\msa@6B
\mathchardef\bumpeq="3\msa@6C
\mathchardef\Bumpeq="3\msa@6D
\mathchardef\lll="3\msa@6E

\mathchardef\ggg="3\msa@6F

\mathchardef\circledS="0\msa@73
\mathchardef\pitchfork="3\msa@74
\mathchardef\dotplus="2\msa@75
\mathchardef\backsim="3\msa@76
\mathchardef\backsimeq="3\msa@77
\mathchardef\complement="0\msa@7B
\mathchardef\intercal="2\msa@7C
\mathchardef\circledcirc="2\msa@7D
\mathchardef\circledast="2\msa@7E
\mathchardef\circleddash="2\msa@7F
\def\ulcorner{\delimiter"4\msa@70\msa@70 }
\def\urcorner{\delimiter"5\msa@71\msa@71 }
\def\llcorner{\delimiter"4\msa@78\msa@78 }
\def\lrcorner{\delimiter"5\msa@79\msa@79 }
\def\yen{\mathhexbox\msa@55 }
\def\checkmark{\mathhexbox\msa@58 }
\def\circledR{\mathhexbox\msa@72 }
\def\maltese{\mathhexbox\msa@7A }
\mathchardef\lvertneqq="3\msb@00
\mathchardef\gvertneqq="3\msb@01
\mathchardef\nleq="3\msb@02
\mathchardef\ngeq="3\msb@03
\mathchardef\nless="3\msb@04
\mathchardef\ngtr="3\msb@05
\mathchardef\nprec="3\msb@06
\mathchardef\nsucc="3\msb@07
\mathchardef\lneqq="3\msb@08
\mathchardef\gneqq="3\msb@09
\mathchardef\nleqslant="3\msb@0A
\mathchardef\ngeqslant="3\msb@0B
\mathchardef\lneq="3\msb@0C
\mathchardef\gneq="3\msb@0D
\mathchardef\npreceq="3\msb@0E
\mathchardef\nsucceq="3\msb@0F
\mathchardef\precnsim="3\msb@10
\mathchardef\succnsim="3\msb@11
\mathchardef\lnsim="3\msb@12
\mathchardef\gnsim="3\msb@13
\mathchardef\nleqq="3\msb@14
\mathchardef\ngeqq="3\msb@15
\mathchardef\precneqq="3\msb@16
\mathchardef\succneqq="3\msb@17
\mathchardef\precnapprox="3\msb@18
\mathchardef\succnapprox="3\msb@19
\mathchardef\lnapprox="3\msb@1A
\mathchardef\gnapprox="3\msb@1B
\mathchardef\nsim="3\msb@1C
\mathchardef\napprox="3\msb@1D
\mathchardef\varsubsetneq="3\msb@20
\mathchardef\varsupsetneq="3\msb@21
\mathchardef\nsubseteqq="3\msb@22
\mathchardef\nsupseteqq="3\msb@23
\mathchardef\subsetneqq="3\msb@24
\mathchardef\supsetneqq="3\msb@25
\mathchardef\varsubsetneqq="3\msb@26
\mathchardef\varsupsetneqq="3\msb@27
\mathchardef\subsetneq="3\msb@28
\mathchardef\supsetneq="3\msb@29
\mathchardef\nsubseteq="3\msb@2A
\mathchardef\nsupseteq="3\msb@2B
\mathchardef\nparallel="3\msb@2C
\mathchardef\nmid="3\msb@2D
\mathchardef\nshortmid="3\msb@2E
\mathchardef\nshortparallel="3\msb@2F
\mathchardef\nvdash="3\msb@30
\mathchardef\nVdash="3\msb@31
\mathchardef\nvDash="3\msb@32
\mathchardef\nVDash="3\msb@33
\mathchardef\ntrianglerighteq="3\msb@34
\mathchardef\ntrianglelefteq="3\msb@35
\mathchardef\ntriangleleft="3\msb@36
\mathchardef\ntriangleright="3\msb@37
\mathchardef\nleftarrow="3\msb@38
\mathchardef\nrightarrow="3\msb@39
\mathchardef\nLeftarrow="3\msb@3A
\mathchardef\nRightarrow="3\msb@3B
\mathchardef\nLeftrightarrow="3\msb@3C
\mathchardef\nleftrightarrow="3\msb@3D
\mathchardef\divideontimes="2\msb@3E
\mathchardef\varnothing="0\msb@3F
\mathchardef\nexists="0\msb@40
\mathchardef\mho="0\msb@66
\mathchardef\thorn="0\msb@67
\mathchardef\beth="0\msb@69
\mathchardef\gimel="0\msb@6A
\mathchardef\daleth="0\msb@6B
\mathchardef\lessdot="3\msb@6C
\mathchardef\gtrdot="3\msb@6D
\mathchardef\ltimes="2\msb@6E
\mathchardef\rtimes="2\msb@6F
\mathchardef\shortmid="3\msb@70
\mathchardef\shortparallel="3\msb@71
\mathchardef\smallsetminus="2\msb@72
\mathchardef\thicksim="3\msb@73
\mathchardef\thickapprox="3\msb@74
\mathchardef\approxeq="3\msb@75
\mathchardef\succapprox="3\msb@76
\mathchardef\precapprox="3\msb@77
\mathchardef\curvearrowleft="3\msb@78
\mathchardef\curvearrowright="3\msb@79
\mathchardef\digamma="0\msb@7A
\mathchardef\varkappa="0\msb@7B
\mathchardef\hslash="0\msb@7D
\mathchardef\hbar="0\msb@7E
\mathchardef\backepsilon="3\msb@7F
\def\Bbb{\ifmmode\let\next\Bbb@\else
 \def\next{\errmessage{Use \string\Bbb\space only in math mode}}\fi\next}
\def\Bbb@#1{{\Bbb@@{#1}}}
\def\Bbb@@#1{\fam\msbfam#1}

\makeatother

\newcommand{\be}{\begin{equation}}
\newcommand{\ee}{\end{equation}}
\newcommand{\bea}{\begin{eqnarray}}
\newcommand{\eea}{\end{eqnarray}}
\newcommand{\nn}{\nonumber}
\newcommand{\ket}[1]{\left| {#1} \right\rangle}
\newcommand{\bra}[1]{\left\langle {#1} \right|}
\newcommand{\spn}[1]{{\rm span}\{{#1}\}}
\newcommand{\binc}[2]{ \left( \begin{array}{c} {#1} \\ {#2}
\end{array} \right) }
\def\bbbz {\Bbb{Z}}

\def\bbbn {\Bbb{N}}
\def\bbbc {\Bbb{C}}
\def\bbbzero {\displaystyle{\Bbb{O}}}
\def\bbbone {{\mathchoice {\rm 1\mskip-4mu l} {\rm 1\mskip-4mu l}
{\rm 1\mskip-4.5mu l} {\rm 1\mskip-5mu l}}}

\title{Singular Vectors of the $N=2$ Superconformal Algebra}
\author{Matthias D\"{o}rrzapf
\thanks{e-mail: M.Doerrzapf@amtp.cam.ac.uk}
\address{Department of Applied Mathematics and Theoretical
Physics\\
University of Cambridge, Silver Street \\
Cambridge, CB3 9EW, U.\ K.\ }}
\pubdate{March 1994}
\pubnumber{DAMTP 94-14 \\ hep-th/9403124}

\abstract{
We derive a set of recursion
formulae to construct singular vectors for the $N=2$ (untwisted)
algebra, by using the approach of Bauer, di Francesco, Itzykson and
Zuber. Applying these
formulae, we obtain explicit expressions for the charged singular
vectors and for a class of uncharged singular vectors.
}

\begin{document}

\maketitle

\section{Introduction}
Recent developments in string theory have shown that the $N=2$
critical string provides a consistent quantum theory of self-dual
gravity in four dimensions \cite{ooguri,nishino1}, and furthermore the
infinite spectrum of particles which appears in other string theories
turns out to be a finite spectrum for $N=2$ critical strings. The interest in
$N=2$ superstring theory is also motivated by the conjecture of
Atiyah and Ward \cite{Atiyah}, that all the bosonic integrable systems
in lower dimensions can be generated by self-dual Yang-Mills theory in
four dimensional space-time with signature $(2,2)$. This leads to the
suggestion that the $N=2$ critical string should give us
insight in integrable systems. Furthermore, the $N=2$
superstring seems to be closely related to the theory of two
dimensional black holes \cite{nishino2}. A treatment of canonical
quantum supergravity in three dimensions and the special r\^{o}le of
the $N=2$ superstring can be found, for instance, in ref. \icite{nicolai}.

It is not only the importance of $N=2$ critical superstring theory
that makes
$N=2$ superconformal theories particularly interesting; there are also
applications in
two dimensional critical phenomena.
Starting with the paper of Belavin, Polyakov and Zamolodchikov
\cite{bpz},
many statistical models at their
critical points have been identified as conformally invariant
\mbox{theories \cite{Friedan2,huse,abf}}.
Furthermore, Friedan, Qiu and Shenker were able to identify the
tricritical Ising model\footnote{Experimentally realised in a ${}^{4}$He
monolayer absorbed on Kr-plated graphite
\cite{ferreira}.} as a $N=1$
superconformal model \cite{Friedan}.
This raised the interesting question if
$N=2$ superconformal theories could be found to describe some
statistical models at their critical points, and indeed it was
shown in ref. \icite{waterson} that under certain circumstances
$O(2)$ Gaussian models are $N=2$ superconformally invariant.

The physically important representations of the $N=2$
superconformal algebra are the irreducible highest weight
representations. In order to understand their structure, it is helpful
to find explicit expressions for the singular vectors appearing in the
Verma modules. Furthermore, since these singular vectors
decouple from the theory, they imply
differential equations for the correlation functions. For the Virasoro
algebra
explicit formulae for some singular vectors
were first given by Benoit and Saint-Aubin \cite{bsa1} and later
for all singular vectors by Bauer et al. \cite{bfiz1,bfiz2} and by
Kent \cite{adrian1}. There are basically three different methods known:
the {\sl ``fusion method'' } used by Bauer et al. \cite{bfiz1,bfiz2},
the {\sl ``analytic continuation method''} developed by Malikov,
Feigin and Fuchs for Kac-Moody algebras \cite{Malikov} and extended by Kent
to the Virasoro algebra \cite{adrian1} and the method of Ganchev and
Petkova which uses the Knizhnik-Zamolodchikov equation to transform a
Kac-Moody singular vector into a Virasoro one \cite{Ganchev}.
The fusion method has been used for various other
chiral algebras as well: for the $N=1$ Neveu-Schwarz algebra by Benoit
and Saint-Aubin \cite{bsa2} and Huang et al. \cite{hwang1}, for the
$N=1$ Ramond algebra by Watts \cite{Gerard}, for the $W_{3}$ algebra by
Bowcock and Watts \cite{Gerard2}, for the affine algebra $A^{(1)}_1$
by Bauer and Sochen \cite{bauer} and for the $W B\!C_{2}$ algebra by
Bajnok \cite{Zoltan}. Recently the analytic continuation method was
used by Bajnok \cite{Zoly} to find $W\!\!A_2$ singular vectors.

We shall show in this paper that the fusion method can be used
successfully in the $N=2$ case. In section 2 we
will introduce the notation and derive the $N=2$ descent equations,
arising from the operator product expansion of $N=2$ superconformal fields.
In the first part of section 3 we use
heuristic arguments to derive a set of recursion formulae which allow us
to compute singular vectors at high levels using known singular vectors
at lower levels. A rigorous proof for these recursion formulae %
is given in the second part of section 3.
In the fourth section we use the recursion formulae to find
explicit expressions for all charged singular vectors and for a class
of uncharged singular vectors which are the analogues of the Benoit
and Saint-Aubin singular vectors in the Virasoro case. Finally we
investigate the dependence of the recursion on the position of the
fusion point.

\section{$N=2$ Superconformal Theories}

\subsection{Notation}
The algebra of chiral superconformal transformations in $N=2$ superconformal
space \cite{Cohn} is generated
 by the super stress-energy tensor $T(Z_{1})$, where $Z_{1}$
denotes a superpoint $(\! z_{1},\theta_{1,1},\theta_{1,2}\! )$. Superconformal
invariance of the theory determines the operator product expansion
(OPE) or short distance expansion of
$T(Z)$ with itself:
\bea
T(Z_{1}) \: T(Z_{2}) &=& -\frac{c}{12 Z_{12}^{2}} + \left( -
\frac{\theta_{12,1} \theta_{12,2}}{Z_{12}^{2}} + \frac{1}{2}
\frac{\theta_{12,2} D_{2,1} - \theta_{12,1} D_{2,2}}{Z_{12}}
 - \frac{\theta_{12,1} \theta_{12,2}}{Z_{12}} \partial_{z_{2}}
\right) T(Z_{2}) , \nn \\
\eea
where we have used $\theta_{12,i}=\theta_{1,i}-\theta_{2,i}$ and
$Z_{12}=z_{1} -z_{2} -\theta_{1,i} \theta_{2,i}$ as superintegrals of
unity and $D_{i,j}=\frac{\partial}{\partial \theta_{i,j}} + \theta_{i,j}
\frac{\partial}{\partial z}$ as superderivatives. If we do not need to
label the superpoint, we shall just write $\theta_{i}$ and
$D_{i}$.

Expanding $T(Z)$ in its modes allows us to find the symmetry algebra
generators\footnote{We put $Z_{2}=0$ to obtain the
field-state correspondence at the origin and we use $L_{m}$ for
$L_{m}(0)$ etc.}:
\bea
T(Z_{1}) & = & -\theta_{12,1} \theta_{12,2} \sum_{m \in \bbbz} Z_{12}^{-m-2}
L_{m}(Z_{2})-\frac{1}{2} \sum_{r \in \bbbz +\frac{1}{2}}
Z_{12}^{-r-\frac{3}{2}} \theta_{12,2} G_{r}^{1}(Z_{2}) \nn \\
&& +\frac{1}{2} \sum_{r \in \bbbz +\frac{1}{2}}
Z_{12}^{-r-\frac{3}{2}} \theta_{12,1} G_{r}^{2}(Z_{2}) - \frac{1}{2}
i \sum_{m \in \bbbz} T_{m} Z_{12}^{-m-1}(Z_{2})  \\
L_{m}(Z_{2}) & = & \oint \frac{dZ_{1}}{2\pi i} Z_{12}^{m+1} T(Z_{1})
\nn \\
G_{r}^{i}(Z_{2}) & = & 2 \oint \frac{dZ_{1}}{2\pi i} \theta_{12,i}
Z_{12}^{r+\frac{1}{2}}T(Z_{1})  \label{eq:gen} \\
T_{m}(Z_{2}) & = & -2i \oint \frac{dZ_{1}}{2\pi i} \theta_{12,1}
\theta_{12,2} Z_{12}^{m} T(Z_{1}) . \nn
\eea
If we use complex co-ordinates for the odd generators
\bea
 \theta^{\pm} &=& \frac{1}{\sqrt{2}} (\theta_{1} \pm i \theta_{2}) \\
G^{\pm} &=& \frac{1}{\sqrt{2}} (G^{1} \pm i G^{2})  \\
 D^{\pm}& =& \frac{1}{\sqrt{2}} (D_{1} \pm i D_{2}) =
\frac{\partial}{\partial \theta^{\mp}} + \theta^{\pm}
\frac{\partial}{\partial z} ,
\eea
we find for the symmetry algebra ${\cal A}$ a
decomposition in two odd fields, the Virasoro algebra and a
$U(1)$ Kac-Moody algebra with the commutation relations:
\bea
[L_{m},L_{n}] & = & (m-n) L_{m+n} + \frac{c}{12} \:(m^{3}-m)\:
\delta_{m+n,0} \nn \\
\ [L_{m},G_{r}^{\pm}] & = & (\frac{1}{2} m-r) G_{m+r}^{\pm} \nn \\
\ [L_{m},T_{n}] & = & -n T_{m+n} \nn \\
\ [T_{m},T_{n}] & = & \frac{1}{3} c m \delta_{m+n,0}  \\
\ [T_{m},G_{r}^{\pm}] & = & \pm G_{m+r}^{\pm} \nn \\
\ \{ G_{r}^{+},G_{s}^{-}\} & = & 2 L_{r+s}+(r-s) T_{r+s} +\frac{c}{3}
(r^{2}-\frac{1}{4}) \delta_{r+s,0} \nn \\
\ \{G_{r}^{+},G_{s}^{+}\} & = & \{G_{r}^{-},G_{s}^{-}\}=0 , \;\;\;\;
\;\;\;\; m,n \in \bbbz ; \: r,s \in \bbbz+\frac{1}{2}. \nn
\eea
The primary fields $\Phi (Z)$ in the theory are those transforming
homogeneously
under superconformal transformations. This leads to a short
distance expansion of $T(Z)$ and $\Phi (Z)$ of the form:
\bea
T(Z_{1}) \Phi (Z_{2}) & = & - \frac{h \theta_{12,1}
\theta_{12,2}}{Z_{12}^{2}} \Phi(Z_{2})+\frac{1}{2}
\frac{\theta_{12,2} D_{2,1} - \theta_{12,1} D_{2,2}}{Z_{12}}
\Phi(Z_{2}) \nn \\
&& -\frac{\theta_{12,1} \theta_{12,2}}{Z_{12}}
\partial_{z_{2}} \Phi(Z_{2}) - \frac{q}{2 Z_{12}} i \Phi (Z_{2}) .
\eea
We call $h$ the {\sl conformal weight} of $\Phi$ and $q$ its {\sl
conformal charge}, corresponding to the scaling dimensions of $L_{0}$
and $T_{0}$ transformations respectively. Using contour integral methods
we obtain the infinitesimal transformations for all generators:
\bea
\ [L_{m},\Phi (Z)] &=& \Bigl\{ h(m+1) z^{m} + \frac{1}{2} (m+1) z^{m}
(\theta^{+} D^{-} + \theta^{-} D^{+}) +z^{m+1} \partial_{z} \nn \\
&& +\frac{q}{2} \theta^{+} \theta^{-} z^{m-1} m (m+1) \Bigr\}
\Phi(Z) \nn \\
\ [G_{r}^{\pm},\Phi (Z)] &=& \Bigl\{ 2 h (r+\frac{1}{2}) \theta^{\pm}
z^{r-\frac{1}{2}} -z^{r+\frac{1}{2}} D^{\pm} \pm \theta^{+} \theta^{-}
(r+\frac{1}{2}) z^{r-\frac{1}{2}} D^{\pm} \nn \\
&& + 2 \theta^{\pm} z^{r+\frac{1}{2}} \partial_{z} \pm q
\theta^{\pm} z^{r-\frac{1}{2}} (r+\frac{1}{2}) \Bigr\} \Phi(Z) \nn \\
\ [T_{m},\Phi(Z)] &=& \Bigl\{ 2 h \theta^{+} \theta^{-} m z^{m-1} +
z^{m} (\theta^{-} D^{+} - \theta^{+} D^{-}) + 2 \theta^{+} \theta^{-}
z^{m} \partial_{z} \nn \\
&& + q z^{m} \Bigr\} \Phi(Z) . \label{eq:phi_cr}
\eea
We should note at this stage that the $N=2$ superconformal algebra we
consider is known as the $N=2$ Neveu-Schwarz or antiperiodic algebra.
In ref. \icite{schwimmer} it has been shown that it is isomorphic
to the $N=2$ Ramond
or periodic algebra which makes a separate discussion redundant. The
$N=2$ twisted algebra does not fit in this framework since the
corresponding fields are not $N=2$ superconformal in our sense,
however a discussion of its singular vectors was given by Semikhatov
\cite{Semikhatov}.
A definition of $N=2$ superconformal theories can be found for
instance in refs. \icite{Kiritsis,Mussardo}.

${\cal A}$ can be decomposed in the usual way:
${\cal A}={\cal
A}_{-} \oplus {\cal H} \oplus {\cal A}_{+}$,
 where
${\cal H}=\spn{L_{0},T_{0}}$ is the
Cartan subalgebra, and\footnote{We write $\bbbn$ for
$\{1,2,3,\ldots\}$ and $\bbbn_{0}$ for $\{0,1,2,\ldots\}$.}
${\cal A}_{\pm}=\spn{ L_{\pm n},T_{\pm n},
G_{\pm r}^{+},G_{\pm r}^{-}: n
\in \bbbn, r \in \bbbn_{0}+\frac{1}{2} }$.
A highest weight vector $\ket{h,q}$ is a simultaneous eigenvector
of ${\cal H}$ with $L_{0}$ and $T_{0}$ eigenvalues $h$ and $q$ respectively,
and ${\cal A}_{+} \ket{h,q}=0$. It is easy to see that a
primary field $\Phi_{h,q}$ generates a highest weight vector
$\ket{h,q}$ on the vacuum: $\ket{h,q}=\Phi_{h,q}(0) \ket{0}$.
If $U({\cal A})$ denotes the universal enveloping algebra of ${\cal
A}$ then the Verma module ${\cal V}_{h,q}$ is defined as ${\cal
V}_{h,q} = U({\cal A}) \otimes_{{\cal H} \oplus {\cal A}_{+}}
\ket{h,q}$. This is the representation of ${\cal A}$ with the basis
\bea
&& \Bigl\{ L_{-i_{I}} \ldots L_{-i_{1}} G^{+}_{-j^{+}_{J^{+}}} \ldots
G^{+}_{-j^{+}_{1}} G^{-}_{-j^{-}_{J^{-}}} \ldots
G^{-}_{-j^{-}_{1}} T_{-k_{K}} \ldots T_{-k_{1}} \ket{h,q} ;
\nn \\
&& i_{I} \geq
\ldots \geq i_{1} \geq 1,\: j^{+}_{J^{+}} > \ldots > j^{+}_{1}
\geq \frac{1}{2} ,\: j^{-}_{J^{-}} > \ldots > j^{-}_{1}
\geq \frac{1}{2} ,\:
k_{K} \geq \ldots \geq k_{1} \geq 1 \Bigr\} . \nn
\eea
Vectors in a Verma module which are not multiples of the highest
weight vector but satisfy the highest weight vector conditions are
called {\sl singular vectors}, i.e. $\Psi_{\tilde{h},\tilde{q}} \in
{\cal V}_{h,q}$ is called singular if $L_{0}
\Psi_{\tilde{h},\tilde{q}} =\tilde{h} \Psi_{\tilde{h},\tilde{q}}$,
$T_{0}
\Psi_{\tilde{h},\tilde{q}} =\tilde{q} \Psi_{\tilde{h},\tilde{q}}$ and
${\cal A}_{+} \Psi_{\tilde{h},\tilde{q}} =0$.

\subsection{Commutation of fields and algebra elements} \label{sctn:sigma}
In this subsection we shall define a representation of the $N=2$
algebra which allows us to interchange fields and algebra elements:
\bea
\Phi_{h,q}(Z) X = \sigma(X) \Phi_{h,q}(Z) & , & X \in {\cal A} . \nn
\eea
Using the equations (\ref{eq:phi_cr}) we find:
\bea
\sigma(L_{m}) &=& L_{m}-h(m+1) z^{m} -z^{m+1} \partial_{z} -
\frac{1}{2} (m+1) z^{m} (\theta^{+} D^{-} +\theta^{-} D^{+}) \nn \\
&& -\frac{q}{2} m(m+1) \theta^{+} \theta^{-} z^{m-1} \nn \\
\sigma(G_{r}^{\pm}) &=& G_{r}^{\pm} -2h(r+\frac{1}{2}) \theta^{\pm}
z^{r-\frac{1}{2}} \mp (r+\frac{1}{2}) \theta^{+} \theta^{-}
z^{r-\frac{1}{2}} D^{\pm} \label{eq:sigma} \\
&& + z^{r+\frac{1}{2}} D^{\pm} -2 \theta^{\pm} z^{r+\frac{1}{2}}
\partial_{z} \mp q (r+\frac{1}{2}) z^{r-\frac{1}{2}} \theta^{\pm} \nn
\\
\sigma(T_{m}) &=& T_{m} - 2hm \theta^{+} \theta^{-} z^{m-1} -z^{m}
(\theta^{-} D^{+} - \theta^{+} D^{-}) - 2 \theta^{+} \theta^{-} z^{m}
\partial_{z} -q z^{m} . \nn
\eea

Trivially, this defines a representation on primary fields. However,
it can be shown by direct computation that it defines a representation
in general.

It is convenient to
choose a basis $l_{m},g_{r}^{i}$ and $t_{m}$
for the algebra ${\cal A}$ such that $\sigma$ does not
involve any derivatives on the basis.
In the manner of Friedan et al. \cite{Friedan} we find that
$\sigma$ terms involving
derivatives arise from singular contributions to the OPE. The only
singular terms we should allow are of the form $\frac{\theta_{12,1}
\theta_{12,2}}{Z_{12}}$. A possible choice is:
\bea
l_{m}(Z_{3}) &=& \oint_{\cal C} \frac{dZ_{1}}{2\pi i} Z_{12}^{m}
Z_{13} T(Z_{1}) \nn \\
g_{r}^{i}(Z_{3}) &=& 2 \oint_{\cal C} \frac{dZ_{1}}{2\pi i}
Z_{12}^{r-\frac{1}{2}} Z_{13} T(Z_{1}) \label{eq:little_i} \\
t_{m}(Z_{3}) &=& -2i \oint_{\cal C} \frac{dZ_{1}}{2\pi i}
Z_{12}^{m-1} Z_{13} T(Z_{1}) , \nn
\eea
where $\cal C$ is a super contour about the super points $Z_{2}$ and
$Z_{3}$, and we shall set $Z_{3}=0$.

For $m \in \bbbn$ and $r \in \bbbn_{0}+\frac{1}{2}$ we shall
use\footnote{It is worth remarking that the $G$'s
and $\theta$'s anticommute.}:
\bea
\begin{array}{cc}
 l_{m}=L_{m}-z L_{m-1} + \frac{1}{2} (\theta^{-}
G_{m-\frac{1}{2}}^{+} + \theta^{+} G_{m-\frac{1}{2}}^{-}) &
\sigma(l_{m})=l_{m}-h z^{m} \\
 g_{r}^{\pm} = G_{r}^{\pm} - z G_{r-1}^{\pm} \mp
\theta^{\pm} T_{r-\frac{1}{2}} & \sigma(g_{r}^{\pm})=g_{r}^{\pm}
-2h \theta^{\pm} z^{r-\frac{1}{2}} \\
 t_{m} = T_{m}- z T_{m-1} & \sigma(t_{m})=t_{m}-2h
\theta^{+} \theta^{-} z^{m-1} , \end{array}  \nn \\ \label{eq:littleg}
\eea
whilst for $-m \in \bbbn$ and $-r \in \bbbn_{0}+\frac{1}{2}$ we use:
\bea
\begin{array}{cc}
 l_{m}=-L_{m}+\frac{1}{z} L_{m+1} + \frac{1}{2z} (\theta^{-}
G_{m+\frac{1}{2}}^{+} + \theta^{+} G_{m+\frac{1}{2}}^{-}) &
\sigma(l_{m})=l_{m}-h z^{m} \\
 g_{r}^{\pm} = -G_{r}^{\pm} +\frac{1}{z} G_{r+1}^{\pm} \mp \frac{1}{z}
\theta^{\pm} T_{r+\frac{1}{2}} & \sigma(g_{r}^{\pm})=g_{r}^{\pm}
-2h \theta^{\pm} z^{r-\frac{1}{2}} \\
 t_{m} = -T_{m}+\frac{1}{z} T_{m+1} & \sigma(t_{m})=t_{m}-2h
\theta^{+} \theta^{-} z^{m-1} , \end{array}  \nn \\ \label{eq:little}
\eea
and finally we have to add $L_{0}$ and $T_{0}$ to obtain a basis.

\subsection{$N=2$ descent equations}
Given an OPE of two primary fields we can derive necessary conditions
for the action of the algebra on the expansion coefficients in the
spirit of Belavin, Polyakov and Zamolodchikov \cite{bpz} who derived
the so called {\sl descent equations} for
the Virasoro case. We start from the most general expression for the
OPE multiplied by the vacuum $\ket{0}$
\bea
\Phi_{h_{1},q_{1}}(Z) \: \Phi_{h_{2},q_{2}}(0) \ket{0} &=&
 \sum_{j \in {\cal J} } C_{12}^{j}
z^{h_{j}-h_{1}-h_{2}} \ket{\psi_{j}(Z)} \nn \\
&& + \theta^{+} \sum_{j \in {\cal J} } C_{12}^{j-}
  z^{h_{j}-h_{1}-h_{2}-\frac{1}{2}} \ket{\psi_{j}^{-}(Z)} \nn \\
 && + \theta^{-} \sum_{j \in {\cal J} } C_{12}^{j+}
 z^{h_{j}-h_{1}-h_{2}-\frac{1}{2}} \ket{\psi_{j}^{+}(Z)} \nn \\
&& + \theta^{+} \theta^{-} \sum_{j \in {\cal J} } \overline{C}_{12}^{j}
 z^{h_{j}-h_{1}-h_{2}-1} \ket{\overline{\psi}_{j}(Z)} ,
\label{eq:ope}
\eea
where $j$ runs through all the conformal families.

The {\sl conformal families} $\ket{\psi_{j}(Z)}$,
$\ket{\psi_{j}^{-}(Z)}$, $\ket{\psi_{j}^{+}(Z)}$ and
$\ket{\overline{\psi}_{j}(Z)}$ can be written as power series in z
\bea
\ket{\psi_{j}(Z)} &=& \sum_{n \in \bbbn_{0}} z^{n} \ket{h_{j}+n,q_{j}} +
\theta^{+} \sum_{r\in \bbbn_{0} +\frac{1}{2}} z^{r-\frac{1}{2}}
\ket{h_{j}+r,q_{j}-1} \label{eq:conf_fam} \\
&& + \theta^{-} \sum_{r\in \bbbn_{0} +\frac{1}{2}} z^{r-\frac{1}{2}}
\ket{h_{j}+r,q_{j}+1} +
\theta^{+} \theta^{-} \sum_{n\in \bbbn_{0}} z^{n-1}
\overline{\ket{h_{j}+n,q_{j}}} , \nn
\eea
where $\ket{h_{j},q_{j}}$ denotes the highest weight vector and
$\ket{h_{j}+n,q_{j}(\pm 1)}$ its descendants;
similarly for $\ket{\psi_{j}^{-}(Z)}$, $\ket{\psi_{j}^{+}(Z)}$ and
$\ket{\overline{\psi}_{j}(Z)}$.

We apply the basis elements of $\cal A$ to equation (\ref{eq:ope}), use
the commutation relations (\ref{eq:phi_cr}) and take in account that
$\Phi_{h_{2},q_{2}}(0)\:\ket{0}$ is a highest weight state. After
performing the differentiations and comparing the coefficients we find
necessary conditions for the expansion coefficients in
(\ref{eq:conf_fam}). We notice that the coefficients
$C_{12}^{j-}$, $C_{12}^{j+}$ and $\overline{C}_{12}^{j}$ have to vanish
necessarily, whilst the $C_{12}^{j}$ can be non trivial if
$q_{j}=q_{1}+q_{2}$. In this case we find the descent equations to be:
\bea
\begin{array}{lcl}
 L_{m} \: \ket{h_{j}+n+m,q_{j}} &=& (mh_{1}+h_{j}-h_{2}+n)
\ket{h_{j}+n,q_{j}} \\[.8mm]
 L_{m} \: \ket{h_{j}+r+m,q_{j}\pm 1} &=&
(h_{1}m+h_{j}-h_{2}+\frac{m}{2}+r) \ket{h_{j}+r,q_{j}\pm 1} \\[.8mm]
 L_{m} \: \overline{\ket{h_{j}+n+m,q_{j}}} &=& (h_{1}m+h_{j}-h_{2}+m+n)
\overline{\ket{h_{j}+n,q_{j}}} \\[.8mm]
&&   + \frac{q_{1}}{2} m(m+1) \ket{h_{j}+n,q_{j}} \\[.8mm]
G_{s}^{+} \: \ket{h_{j}+r+s,q_{j}} &=& - \ket{h_{j}+r,q_{j}+1} \\[.8mm]
G_{s}^{+} \: \ket{h_{j}+n+s,q_{j}-1} &=&
-(2h_{1}s+h_{j}-h_{2}+n+q_{1}(s+\frac{1}{2})) \ket{h_{j}+n,q_{j}} \\[.8mm]
&&  - \overline{\ket{h_{j}+n,q_{j}}} \\[.8mm]
 G_{s}^{+} \: \ket{h_{j}+n+s,q_{j}+1} &=& 0 \\[.8mm]
 G_{s}^{+} \: \overline{\ket{h_{j}+r+s,q_{j}}} &=& (2h_{1}s +h_{j}
-h_{2} +r+s+q_{1}(s+\frac{1}{2})) \ket{h_{j}+r,q_{j}+1} \\[.8mm]
 G_{s}^{-} \: \ket{h_{j}+r+s,q_{j}} &=& - \ket{h_{j}+r,q_{j}-1} \\[.8mm]
 G_{s}^{-} \: \ket{h_{j}+n+s,q_{j}-1} &=& 0 \\[.8mm]
 G_{s}^{-} \: \ket{h_{j}+n+s,q_{j}+1} &=&
-(2h_{1}s+h_{j}-h_{2}+n-q_{1}(s+\frac{1}{2}) \ket{h_{j}+n,q_{j}} \\[.8mm]
&& + \overline{\ket{h_{j}+n,q_{j}}} \\[.8mm]
 G_{s}^{-} \: \overline{\ket{h_{j}+r+s,q_{j}}} &=& - (2h_{1}s +h_{j}
-h_{2} +r+s-q_{1}(s+\frac{1}{2})) \ket{h_{j}+r,q_{j}-1}  \\[.8mm]
 T_{m} \: \ket{h_{j}+n+m,q_{j}} &=& q_{1} \ket{h_{j}+n,q_{j}} \\[.8mm]
 T_{m} \: \ket{h_{j}+r+m,q_{j}\pm 1} &=&
(\pm 1 +q_{1}) \ket{h_{j}+r,q_{j}\pm 1} \\[.8mm]
 T_{m} \: \overline{\ket{h_{j}+n+m,q_{j}}} &=& 2 h_{1}m
\ket{h_{j}+n,q_{j}} + q_{1} \overline{\ket{h_{j}+n,q_{j}}}  \\[.8mm]
 L_{0} \: \ket{h_{j}+n,q_{j}} &=& (h_{j}+n) \ket{h_{j}+n,q_{j}} \\[.8mm]
 L_{0} \: \ket{h_{j}+r,q_{j}\pm 1} &=&
(h_{j}+r) \ket{h_{j}+r,q_{j}\pm 1} \\[.8mm]
 L_{0} \: \overline{\ket{h_{j}+n,q_{j}}} &=& (h_{j}+n)
\overline{\ket{h_{j}+n,q_{j}}} \\[.8mm]
 T_{0} \: \ket{h_{j}+n,q_{j}} &=& q_{j} \ket{h_{j}+n,q_{j}} \\[.8mm]
 T_{0} \: \ket{h_{j}+r,q_{j}\pm 1} &=&
(q_{j} \pm 1) \ket{h_{j}+r,q_{j}\pm 1} \\[.8mm]
 T_{0} \: \overline{\ket{h_{j}+n,q_{j}}} &=& q_{j}
\overline{\ket{h_{j}+n,q_{j}}}\: ,
\label{eq:descent}
\end{array}
\eea
where we have put $q_{j}=q_{1}+q_{2}$ and $m \in \bbbn$, $n \in
\bbbn_{0}$, $r,s \in \bbbn_{0}+\frac{1}{2}$.

In theory, we only have to solve this system of infinitely many
equations for infinitely many unknowns to find out which conformal
families contribute to the right hand side of
equation (\ref{eq:ope}) and this would determine the OPE completely.
However, this seems impractical and we have to rely on other methods to
determine the {\sl fusion rules} \cite{Kiritsis,Mussardo,Matthias}.

\subsection{The $N=2$ determinant formula}
The determinant formula for the $N=2$ superconformal (antiperiodic)
algebra was
first given by Boucher, Friedan and Kent \cite{Adrian}. If we use the
parametrisation:
\bea
c &=& 3-3t  \nn \\
h_{r,s} &=& \frac{r^{2}-1}{8} t - \frac{rs}{4} + \frac{s^{2}-1}{8t}
 - \frac{4 q^{2}-1}{8t} \\
h_{k} &=& k q +\frac{1}{2} t (k^{2}-\frac{1}{4}) , \nn
\eea
where $r$ is chosen to be $\in \bbbz$, $s \in 2\bbbz$ and $k \in
\bbbz+\frac{1}{2}$, then the determinant formula can be written as
\bea
{\rm det}(M_{n,m}(c,h,q)) \propto
\displaystyle{ \prod_{ 1 \leq rs \leq
2n \atop  r \in \bbbn \; , \; s \in 2 \bbbn}}
(h-h_{r,s})^{p(n-\frac{rs}{2},m)} \prod_{k \in \bbbz+\frac{1}{2}}
(h-h_{k})^{\tilde{p}(n-|k|,m-sgn(k),k)} ,
\eea
where $p(n-\frac{rs}{2},m)$ and $\tilde{p}(n-|k|,m-sgn(k),k)$ are the
corresponding partition functions. This makes it apparent that
for every $t$ and $q$, every positive integer $r$ and every positive even $s$
the Verma module ${\cal V}_{c(t),h_{r,s}(t,q),q}$ has
an uncharged singular vector at level $\frac{rs}{2}$.
Similarly the Verma module ${\cal V}_{c(t),h_{k}(t,q),q}$ has a
singular vector at level $|k|$ with charge ${\rm sgn}(k)$.
By direct computation we can find for
example the singular vector of the Verma module
${\cal V}_{c(t),h_{1,2}(t,q),q}$:
\bea
\Psi_{1,2} = \left( t L_{-1} + \frac{t}{q-1}
G_{-\frac{1}{2}}^{+} G_{-\frac{1}{2}}^{-} + (q+1) T_{-1} \right)
\ket{h_{1,2},q}  . \label{eq:svec_12}
\eea
We will use this singular vector in section 4 to compute singular vectors at
higher levels recursively.

\section{Recursion formulae for singular vectors}

\subsection{Recursion formulae} \label{sctn:rec}
The following arguments should be understood on a heuristic level
only,
in order to derive the recursion formulae. In the next section
we will prove these recursion formulae rigorously without using any of
the heuristic arguments.

Let us consider a three point function of primary fields
\bea
\bra{h_{j},q_{j}} \Phi_{h_{1},q_{1}}(Z) \Phi_{h_{2},q_{2}}(0) \ket{0} ,
\label{eq:three_pf}
\eea
where $\ket{h_{j},q_{j}}$ is a highest weight vector generated by a
primary field $\Phi_{h_{j},q_{j}}$.
(\ref{eq:three_pf}) will only be non trivial if and only if the
descent equations (\ref{eq:descent})
allow a non trivial contribution of the highest weight state
$\ket{h_{j},q_{j}}$ to the right hand side of the OPE (\ref{eq:ope}).
In this case we say we have an allowed {\sl (even) fusion} of
$\Phi_{h_{1},q_{1}}$ and $\Phi_{h_{2},q_{2}}$ to $\Phi_{h_{j},q_{j}}$.
The information which three point functions are non trivial is
therefore summarised in so called {\sl fusion rules}. For the $N=2$
case a discussion of them\footnote{As far as we know, the descent
equations
decouple for even and odd contributions and we can consider ``even''
and ``odd'' fusion separately. We will not need to use odd fusion and
we write fusion for even fusion below.}
can be found in refs. \icite{Matthias,Kiritsis,Mussardo}. Let us
furthermore assume (\ref{eq:three_pf}) is non trivial which means we
pick an allowed fusion and suppose we know a singular vector in the
Verma module with highest weight state $\Phi_{h_{2},q_{2}}(0) \ket{0}$,
$\Psi$ say, at level $N$:
\bea
\Psi={\cal N} \Phi_{h_{2},q_{2}}(0) \ket{0} ,
\eea
where ${\cal N}$ is a linear combination of algebra generators $\in
{\cal A}_{-}$.

Since the singular vectors decouple, the
function $\bra{h_{j},q_{j}} \Phi_{h_{1},q_{1}}(Z) {\cal{N}}
\Phi_{h_{2},q_{2}}(0) \ket{0}$ has to vanish identically\footnote{Recall
that $\Psi$ is one representative of the zero vector in the irreducible
representation with highest weight state $\ket{h_{2},q_{2}}$
.}. We use at this stage the representation $\sigma$
defined in section \ref{sctn:sigma} in order to commute ${\cal N}$ and
$\Phi_{h_{1},q_{1}}(Z)$:
\bea
0 = \bra{h_{j},q_{j}} \Phi_{h_{1},q_{1}}(Z) {\cal{N}}
\Phi_{h_{2},q_{2}}(0) \ket{0} = \bra{h_{j},q_{j}} \sigma({\cal N})
\Phi_{h_{1},q_{1}}(Z) \Phi_{h_{2},q_{2}}(0) \ket{0} . \label{eq:null}
\eea
We write ${\cal N}$ in the basis
(\ref{eq:little}) before applying $\sigma$ to it. This avoids any
derivatives\footnote{However, we have to deal with the
superderivatives now on an algebraic level in form of the descent
equations.} in $\sigma({\cal N})$.

Let $\Pi_{j}$ be the projection operator onto the conformal family
generated by the
primary field $\Phi_{h_{j},q_{j}}$. Then define the following OPEs:
\bea
{\cal{F}}(Z) &=& \Pi_{j} (\Phi_{h_{1},q_{1}}(Z) \Phi_{h_{2},q_{2}}(0)
\ket{0}) =
\sum_{n \geq 0} z^{h_{j}-h_{1}-h_{2}+n} f_{n}
+ \theta^{+} \sum_{r \in \bbbn_{0} +\frac{1}{2}}
z^{h_{j}-h_{1}-h_{2}+r-\frac{1}{2}} f_{r}^{-} \nn \\
&&  + \theta^{-} \sum_{ r \in \bbbn_{0} +\frac{1}{2}}
z^{h_{j}-h_{1}-h_{2}+r-\frac{1}{2}} f_{r}^{+} +
\theta^{+} \theta^{-} \sum_{n \geq 0} z^{h_{j}-h_{1}-h_{2}+n-1}
\overline{f}_{n} \label{eq:fz} \\
{\cal J}(Z) &=& \sigma({\cal N}) {\cal F}(Z) =
\sum_{n \geq 0} z^{h_{j}-h_{1}-h_{2}+n-N} j_{n}
+ \theta^{+} \sum_{ r \in \bbbn_{0} +\frac{1}{2}}
z^{h_{j}-h_{1}-h_{2}+r-\frac{1}{2}-N} j_{r}^{-} \nn \\
&&  + \theta^{-} \sum_{ r \in \bbbn_{0} +\frac{1}{2}}
z^{h_{j}-h_{1}-h_{2}+r-\frac{1}{2}-N} j_{r}^{+} +
\theta^{+} \theta^{-} \sum_{n \geq 0} z^{h_{j}-h_{1}-h_{2}+n-1-N}
\overline{j}_{n} \label{eq:jz}
\eea
Since we are considering an allowed fusion, ${\cal F}(Z)$ is non
trivial. However, according to (\ref{eq:null}) we
expect ${\cal J}(Z)$
to decouple. This means that it lies completely in the submodule which we
have to quotient out of the Verma module ${\cal V}_{h_{j},q_{j}}$ in
order to get the corresponding irreducible representation. The
expansion coefficients of ${\cal J}(Z)$ must lie in the submodule
generated by a singular vector.
Hence, at the level where we
expect the first singular vector in ${\cal V}_{h_{j},q_{j}}$ the coefficients
of ${\cal J}(Z)$ have to vanish or give us an expression for the
singular vector.
In the following we derive recursion formulae for
the coefficients of ${\cal F}(Z)$ in order to set consistently the
coefficients of ${\cal J}(Z)$ equal to 0. We will see that in
certain cases this recursion procedure breaks down at the level where
we expect the first singular vector in ${\cal V}_{h_{j},q_{j}}$ and the
information achieved from the recursion for ${\cal F}(Z)$ will allow
us to compute the coefficients of ${\cal J}(Z)$ for this level and
hence the singular vector.

We write $\sigma({\cal N})$ again in the standard basis:
\bea
\sigma({\cal N}) &=& \sum_{k=0}^{N} z^{-N+k} S_{k} \nn \\
&& + \sum_{i=0}^{N-1} \sum_{k=0}^{N} z^{-N+k-i-1}
(p_{k}^{1,i} L_{1} +
p_{k}^{2,i} G_{\frac{1}{2}}^{+} G_{\frac{1}{2}}^{-} +
p_{k-\frac{1}{2}}^{3^{-},i} G_{\frac{1}{2}}^{+} +
p_{k-\frac{1}{2}}^{4^{+},i} G_{\frac{1}{2}}^{-} )
L_{1}^{i}
\nn \\
&& + \theta^{+} \sum_{j=\frac{1}{2}}^{N-\frac{1}{2}} z^{-N+j-\frac{1}{2}}
S_{j}^{-} \nn \\
&& + \theta^{+} \sum_{i=0}^{N-1} \sum_{j=\frac{1}{2}}^{N+\frac{1}{2}}
 z^{-N+j-i-\frac{3}{2}}
(p_{j}^{-,1^{-},i} L_{1} +
p_{j}^{-,2^{-},i} G_{\frac{1}{2}}^{+} G_{\frac{1}{2}}^{-} +
p_{j-\frac{1}{2}}^{-,3^{--},i} G_{\frac{1}{2}}^{+} +
p_{j-\frac{1}{2}}^{-,4,i} G_{\frac{1}{2}}^{-} )
L_{1}^{i}
\nn \\
&& + \theta^{-} \sum_{j=\frac{1}{2}}^{N-\frac{1}{2}} z^{-N+j-\frac{1}{2}}
S_{j}^{+} \nn \\
&& + \theta^{-} \sum_{i=0}^{N-1} \sum_{j=\frac{1}{2}}^{N+\frac{1}{2}}
 z^{-N+j-i-\frac{3}{2}} (p_{j}^{+,1^{+},i} L_{1} +
p_{j}^{+,2^{+},i} G_{\frac{1}{2}}^{+} G_{\frac{1}{2}}^{-} +
p_{j-\frac{1}{2}}^{+,3,i} G_{\frac{1}{2}}^{+} +
p_{j-\frac{1}{2}}^{+,4^{++},i} G_{\frac{1}{2}}^{-} )
L_{1}^{i}
\nn \\
&& + \theta^{+} \theta^{-} \sum_{k=0}^{N} z^{-N+k-1} \overline{S}_{k} \nn \\
&& + \theta^{+} \theta^{-}
\sum_{i=0}^{N-1} \sum_{k=0}^{N} z^{-N+k-i-2}
(\overline{p}_{k}^{1,i} L_{1} +
\overline{p}_{k}^{2,i} G_{\frac{1}{2}}^{+} G_{\frac{1}{2}}^{-} +
\overline{p}_{k-\frac{1}{2}}^{3^{-},i} G_{\frac{1}{2}}^{+} +
\overline{p}_{k-\frac{1}{2}}^{4^{+},i} G_{\frac{1}{2}}^{-} )
L_{1}^{i}
 \nn, \\
\label{eq:sz} \eea
\newcommand{\Pki}{(p_{m}^{1,i} L_{1} +
p_{m}^{2,i} G_{\frac{1}{2}}^{+} G_{\frac{1}{2}}^{-} +
p_{m-\frac{1}{2}}^{3^{-},i} G_{\frac{1}{2}}^{+} +
p_{m-\frac{1}{2}}^{4^{+},i} G_{\frac{1}{2}}^{-} )
L_{1}^{i}}
\newcommand{\Ppki}{(p_{j}^{+,1^{+},i} L_{1} +
p_{j}^{+,2^{+},i} G_{\frac{1}{2}}^{+} G_{\frac{1}{2}}^{-} +
p_{j-\frac{1}{2}}^{+,3,i} G_{\frac{1}{2}}^{+} +
p_{j-\frac{1}{2}}^{+,4^{++},i} G_{\frac{1}{2}}^{-} )
L_{1}^{i}}
\newcommand{\Pmki}{(p_{j}^{-,1^{-},i} L_{1} +
p_{j}^{-,2^{-},i} G_{\frac{1}{2}}^{+} G_{\frac{1}{2}}^{-} +
p_{j-\frac{1}{2}}^{-,3^{--},i} G_{\frac{1}{2}}^{+} +
p_{j-\frac{1}{2}}^{-,4,i} G_{\frac{1}{2}}^{-} )
L_{1}^{i}}
\newcommand{\Pbki}{
(\overline{p}_{m}^{1,i} L_{1} +
\overline{p}_{m}^{2,i} G_{\frac{1}{2}}^{+} G_{\frac{1}{2}}^{-} +
\overline{p}_{m-\frac{1}{2}}^{3^{-},i} G_{\frac{1}{2}}^{+} +
\overline{p}_{m-\frac{1}{2}}^{4^{+},i} G_{\frac{1}{2}}^{-} )
L_{1}^{i}}
where the coefficients\footnote{The index coincides with the level of
the coefficient.} $S_{k},S^{-}_{j},S^{+}_{j},\overline{S}_{k}$
and $p_{k},p^{-}_{j},p^{+}_{j},
\overline{p}_{k}$ only contain algebra elements in ${\cal A}_{-} \oplus
{\cal H}$. This allows us to write the coefficients $j_{n},j_{r}^{+},
j_{r}^{-}$ and $\overline{j}_{n}$
of ${\cal J}(Z)$ in terms of
$f_{n},f_{r}^{+},f_{r}^{-}$ and $\overline{f}_{n}$ using equations
(\ref{eq:fz},\ref{eq:jz},\ref{eq:sz}).
\bea
j_{n} \!\!\!\! &=& \!\!\!  \sum_{m=0}^{n} S_{m} f_{n-m}
+ \sum_{i=0}^{N-1} \sum_{m=0}^{n} \Pki f_{n-m+i+1} \nn \\
\overline{j}_{n}\!\!\!\! &=& \!\!\! \sum_{m=0}^{n} S_{m} \overline{f}_{n-m}
+ \sum_{i=0}^{N-1} \sum_{m=0}^{n} \Pki \overline{f}_{n-m+i+1} \nn
\\
&& \!\!\! + \sum_{m=0}^{n} \overline{S}_{m} f_{n-m}
+ \sum_{i=0}^{N-1} \sum_{m=0}^{n} \Pbki f_{n-m+i+1} \nn
\\
&& \!\!\! - \sum_{j=\frac{1}{2}}^{n-\frac{1}{2}} S_{j}^{-} f_{n-j}^{+}
- \sum_{i=0}^{N-1} \sum_{j=\frac{1}{2}}^{n+\frac{1}{2}} \Pmki
f_{n-j+i+1}^{+} \nn
\\
&& \!\!\! + \sum_{j=\frac{1}{2}}^{n-\frac{1}{2}} S_{j}^{+} f_{n-j}^{-}
+ \sum_{i=0}^{N-1} \sum_{j=\frac{1}{2}}^{n+\frac{1}{2}} \Ppki
f_{n-j+i+1}^{-} \nn
\\
j_{r}^{-}\!\!\!\! &=& \!\!\! \sum_{m=0}^{r-\frac{1}{2}} S_{m} f_{r-m}^{-} +
\sum_{i=0}^{N-1} \sum_{m=0}^{r+\frac{1}{2}} \Pki f_{r-m+i+1}^{-}
\nn \\
&& \!\!\! + \sum_{j=\frac{1}{2}}^{r} S_{j}^{-} f_{r-j} + \sum_{i=0}^{N-1}
\sum_{j=\frac{1}{2}}^{r} \Pmki f_{r-j+i+1} \nn \\
j_{r}^{+}\!\!\!\! &=&\!\!\! \sum_{m=0}^{r-\frac{1}{2}} S_{m} f_{r-m}^{+} +
\sum_{i=0}^{N-1} \sum_{m=0}^{r+\frac{1}{2}} \Pki f_{r-m+i+1}^{+}
\nn \\
&& \!\!\! + \sum_{j=\frac{1}{2}}^{r} S_{j}^{+} f_{r-j} + \sum_{i=0}^{N-1}
\sum_{j=\frac{1}{2}}^{r} \Ppki f_{r-j+i+1} \nn \\ \label{eq:j_f}
\eea
We can use the descent equations (\ref{eq:descent}) to evaluate the
action of $L_{1}$, $G_{\frac{1}{2}}^{+}$ and $G_{\frac{1}{2}}^{-}$ on
$f_{n},f_{r}^{+},f_{r}^{-}$ and $\overline{f}_{n}$:\footnote{We define
the empty product $\prod_{i=m}^{n} \ldots$ for $m>n$ to be $1$, and
the empty sum $\sum_{i=m}^{n} \ldots$ for $m>n$ to vanish.}
\bea
L_{1}^{i+1} f_{m} &=& \prod_{j=0}^{i} (h_{1}+L_{0}-h_{2}+j)
f_{m-i-1} \nn \\
L_{1}^{i+1} f_{r}^{\pm} &=& \prod_{j=0}^{i}
(h_{1}+L_{0}-h_{2}+j+\frac{1}{2}) f_{r-i-1}^{\pm} \nn \\
L_{1}^{i+1} \overline{f}_{m} &=& \prod_{j=0}^{i}
(h_{1}+L_{0}-h_{2}+j+1) \overline{f}_{m-i-1} \nn \\
&& +q_{1} \sum_{l=0}^{i} \prod_{j_{1}=l+1}^{i}
(h_{1}+L_{0}-h_{2}+j_{1}+1) \prod_{j_{2}=0}^{l-1}
(h_{1}+L_{0}-h_{2}+j_{2}) f_{m-i-1} \nn \\
G_{\frac{1}{2}}^{+} L_{1}^{i} f_{m} &=& -\prod_{j=0}^{i-1}
(h_{1}+L_{0}-h_{2}+j+\frac{1}{2}) f_{m-i-\frac{1}{2}}^{+} \nn \\
G_{\frac{1}{2}}^{+} L_{1}^{i} f_{r}^{+} &=& 0 \nn \\
G_{\frac{1}{2}}^{+} L_{1}^{i} f_{r}^{-} &=& -\prod_{j=0}^{i-1}
(h_{1}+L_{0}-h_{2}+j+1) \Bigl( (h_{1}+L_{0}-h_{2}+q_{1})
f_{r-i-\frac{1}{2}} + \overline{f}_{r-i-\frac{1}{2}} \Bigr) \nn \\
G_{\frac{1}{2}}^{+} L_{1}^{i} \overline{f}_{m} &=&
\Bigl( \prod_{j=0}^{i-1} (h_{1}+L_{0}-h_{2}+j+\frac{3}{2})
(h_{1}+L_{0}-h_{2}+\frac{1}{2}+q_{1}) \nn \\
&& - q_{1} \sum_{l=0}^{i-1}
\prod_{j_{1}=l+1}^{i-1} (h_{1}+L_{0}-h_{2}+j_{1}+\frac{3}{2})
\prod_{j_{2}=0}^{l-1} (h_{1}+L_{0}-h_{2}+j_{2}+\frac{1}{2}) \Bigr)
f_{m-i-\frac{1}{2}}^{+} \nn \\
G_{\frac{1}{2}}^{-} L_{1}^{i} f_{m} &=& -\prod_{j=0}^{i-1}
(h_{1}+L_{0}-h_{2}+j+\frac{1}{2}) f_{m-i-\frac{1}{2}}^{-} \nn \\
G_{\frac{1}{2}}^{-} L_{1}^{i} f_{r}^{+} &=& -\prod_{j=0}^{i-1}
(h_{1}+L_{0}-h_{2}+j+1) \Bigl( (h_{1}+L_{0}-h_{2}-q_{1})
f_{r-i-\frac{1}{2}} - \overline{f}_{r-i-\frac{1}{2}} \Bigr) \nn \\
G_{\frac{1}{2}}^{-} L_{1}^{i} f_{r}^{-} &=& 0 \nn \\
G_{\frac{1}{2}}^{-} L_{1}^{i} \overline{f}_{m} &=&
- \Bigl( \prod_{j=0}^{i-1} (h_{1}+L_{0}-h_{2}+j+\frac{3}{2})
(h_{1}+L_{0}-h_{2}+\frac{1}{2}-q_{1}) \nn \\
&& + q_{1} \sum_{l=0}^{i-1}
\prod_{j_{1}=l+1}^{i-1} (h_{1}+L_{0}-h_{2}+j_{1}+\frac{3}{2})
\prod_{j_{2}=0}^{l-1} (h_{1}+L_{0}-h_{2}+j_{2}+\frac{1}{2}) \Bigr)
f_{m-i-\frac{1}{2}}^{-} \nn \\
G_{\frac{1}{2}}^{+} G_{\frac{1}{2}}^{-} L_{1}^{i} f_{m} &=& \prod_{j=0}^{i-1}
(h_{1}+L_{0}-h_{2}+j+1) \Bigl( (h_{1}+L_{0}-h_{2}+q_{1}) f_{m-i-1} +
\overline{f}_{m-i-1} \Bigr) \nn \\
G_{\frac{1}{2}}^{+} G_{\frac{1}{2}}^{-} L_{1}^{i} f_{r}^{+} &=&
 2 \prod_{j=0}^{i-1}
(h_{1}+L_{0}-h_{2}+j+\frac{3}{2})
(h_{1}+L_{0}-h_{2}+\frac{1}{2}) f_{r-i-1}^{+} \nn \\
G_{\frac{1}{2}}^{+} G_{\frac{1}{2}}^{-} L_{1}^{i} f_{r}^{-} &=& 0 \nn \\
G_{\frac{1}{2}}^{+} G_{\frac{1}{2}}^{-} L_{1}^{i} \overline{f}_{m} &=&
 \Bigl(
\prod_{j=0}^{i-1} (h_{1}+L_{0}-h_{2}+j+2) (h_{1}+L_{0}-h_{2}+1-q_{1})
\nn \\ &&
+ q_{1} \sum_{l=0}^{i-1} \prod_{j_{1}=l+1}^{i-1}
(h_{1}+L_{0}-h_{2}+j_{1}+2) \prod_{j_{2}=0}^{l-1}
(h_{1}+L_{0}-h_{2}+j_{2}+1) \Bigr) \nn \\
&& \Bigl( (h_{1}+L_{0}-h_{2}+q_{1})
f_{m-i-1} + \overline{f}_{m-i-1} \Bigr)\: . \label{eq:fdesc}
\eea
 As mentioned above
we set $j_{n},j_{r}^{+},j_{r}^{-}$ and $\overline{j}_{n}$ equal to
0, and solve the equations (\ref{eq:j_f}) for $f_{n},f_{r}^{+},f_{r}^{-}$
and $\overline{f}_{n}$, leading to a set of recursion formulae for them.
The recursion procedure splits up in ``even'' (\ref{eq:step_e})
and ``odd'' (\ref{eq:step_o}) steps which
we have to perform alternately.
\bea
\left( \begin{array}{c}
f_{n} \\ \overline{f}_{n} \end{array} \right) &=&
- {\cal S}_{h_{j}+n,q_{j}}^{-1} \left( \begin{array}{c} Y_{n}^{1} \\ Y_{n}^{2}
\end{array} \right) \label{eq:step_e}
\eea
\bea
f_{r}^{+} = - \frac{1}{s^{+}_{h_{j}+r,q_{j}+1}} Y_{r}^{+} &  &
f_{r}^{-} = - \frac{1}{s^{-}_{h_{j}+r,q_{j}-1}} Y_{r}^{-}  \label{eq:step_o}
\eea
The matrix ${\cal S}$ and the coefficients $s^{+}, s^{-}$
depend on $L_{0}$ and hence on the level $n$ of
$f_{n}, f_{n}^{\pm}$ and $\overline{f}_{n}$. If necessary we shall
assign a subscript denoting the level.
 ${\cal S}$ has to be inverted for each
even step; if this fails at a certain level $m$, we
can not define the corresponding $j_{m}$ and $\overline{j}_{m}$ to be
$0$. Similarly we have to divide by $s^{+}$ and $s^{-}$ for each odd
step which leads to non trivial $j_{r}^{\pm}$ if the division
fails:
\bea
s_{11} &=& S_{0} + \sum_{i=0}^{N-1} \Bigl\{ p_{0}^{1,i} \prod_{j=0}^{i}
(h_{1}+L_{0}-h_{2}+j) +p_{0}^{2,i} \prod_{j=0}^{i-1}
(h_{1}+L_{0}-h_{2}+j+1) \nn \\
&& (h_{1}+L_{0}-h_{2}+q_{1}) \Bigr\} \nn \\
s_{12} &=& \sum_{i=0}^{N-1} p_{0}^{2,i}
\prod_{j=0}^{i-1} (h_{1}+L_{0}-h_{2}+j+1)
\nn \\
s_{21} &=& \overline{S}_{0} + \sum_{i=0}^{N-1} \Bigl\{ q_{1}
p_{0}^{1,i} \sum_{l=0}^{i}
\prod_{j_{1}=l+1}^{i}(h_{1}+L_{0}-h_{2}+j_{1}+1) \prod_{j_{2}=0}^{l-1}
(h_{1}+L_{0}-h_{2}+j_{2}) \nn \\
&& + p_{0}^{2,i} \Bigl( \prod_{j=0}^{i-1}
(h_{1}+L_{0}-h_{2}+j+2) (h_{1}+L_{0}-h_{2}+1-q_{1}) + q_{1}
\nn \\
&&
\sum_{l=0}^{i-1} \prod_{j_{1}=l+1}^{i-1} (h_{1}+L_{0}-h_{2}+j_{1}+2)
\prod_{j_{2}=0}^{l-1} (h_{1}+L_{0}-h_{2}+j_{2}+1)
\Bigr) (h_{1}+L_{0}-h_{2}+q_{1}) \nn \\
&& + \overline{p}_{0}^{1,i} \prod_{j=0}^{i} (h_{1}+L_{0}-h_{2}+j)
+ \overline{p}_{0}^{2,i} \prod_{j=0}^{i-1} (h_{1}+L_{0}-h_{2}+j+1)
(h_{1}+L_{0}-h_{2}+q_{1}) \nn \\
&& + \prod_{j=0}^{i-1} (h_{1}+L_{0}-h_{2}+j+1)
( p_{0}^{-,4,i}(h_{1}+L_{0}-h_{2}-q_{1}) -p_{0}^{+,3,i}
(h_{1}+L_{0}-h_{2}+q_{1}))  \Bigr\} \nn \\
s_{22} &=& S_{0} + \sum_{i=0}^{N-1} \Bigl\{
p_{0}^{1,i} \prod_{j=0}^{i} (h_{1}+L_{0}-h_{2}+j+1)
+p_{0}^{2,i} \Bigl( \prod_{j=0}^{i-1}
(h_{1}+L_{0}-h_{2}+j+2) \nn \\
&& (h_{1}+L_{0}-h_{2}+1-q_{1}) + q_{1}
\sum_{l=0}^{i-1} \prod_{j_{1}=l+1}^{i-1} (h_{1}+L_{0}-h_{2}+j_{1}+2)
\nn
\\  && \prod_{j_{2}=0}^{l-1} (h_{1}+L_{0}-h_{2}+j_{2}+1)
\Bigr)
+ \overline{p}_{0}^{2,i} \prod_{j=0}^{i-1} (h_{1}+L_{0}-h_{2}+j+1)
\nn \\ &&
- \prod_{j=0}^{i-1} (h_{1}+L_{0}-h_{2}+j+1)
(p_{0}^{-,4,i}+p_{0}^{+,3,i}) \Bigr\} \label{eq:s_coef}
\eea
\bea
{\cal S} &=& \left( \begin{array}{cc} s_{11} & s_{12} \\ s_{21} &
s_{22} \end{array} \right)  \label{eq:rec_s}
\eea
\bea
s^{+} &=&
S_{0} + \sum_{i=0}^{N-1} \Bigl\{ p_{0}^{1,i} \prod_{j=0}^{i}
(h_{1}+L_{0}-h_{2}+j+\frac{1}{2}) + 2 p_{0}^{2,i} \prod_{j=0}^{i-1}
(h_{1}+L_{0}-h_{2}+j+\frac{3}{2}) \nn \\
&&
(h_{1}+L_{0}-h_{2}+\frac{1}{2}) - p_{0}^{+,3,i}
\prod_{j=0}^{i-1} (h_{1}+L_{0}-h_{2}+j+\frac{1}{2}) \Bigr\} \nn \\
s^{-} &=&
S_{0} + \sum_{i=0}^{N-1} \Bigl\{ p_{0}^{1,i} \prod_{j=0}^{i}
(h_{1}+L_{0}-h_{2}+j+\frac{1}{2}) - p_{0}^{-,4,i}
\prod_{j=0}^{i-1} (h_{1}+L_{0}-h_{2}+j+\frac{1}{2}) \Bigr\}\: . \nn \\
\eea
The expressions for $Y_{n}^{1}$ and  $Y_{n}^{2}$
depend on $f_{m},f_{s}^{\pm}$ and $\overline{f}_{m}$ $(m,s<n)$ , similarly
for $Y_{r}^{\pm}$, so that (\ref{eq:step_e},\ref{eq:step_o}) really
defines a recursion. For convenience we give the following expressions
for $Y_{n}^{1} , Y_{n}^{2}$ and $Y_{r}^{\pm}$ in which we still have to
insert equations (\ref{eq:fdesc}):
\bea
Y^{1}_{n} \!\!\!\!&=&\!\!\!\! \sum_{m=1}^{n} S_{m} f_{n-m}
+ \sum_{i=0}^{N-1} \sum_{m=1}^{n} \Pki f_{n-m+i+1} \nn \\
Y^{2}_{n} \!\!\!\!&=&\!\!\!\! \sum_{m=1}^{n} S_{m} \overline{f}_{n-m}
+ \sum_{i=0}^{N-1} \sum_{m=1}^{n} \Pki \overline{f}_{n-m+i+1} \nn
\\
&& \!\!\!\! + \sum_{m=1}^{n} \overline{S}_{m} f_{n-m}
+ \sum_{i=0}^{N-1} \sum_{m=1}^{n} \Pbki f_{n-m+i+1} \nn
\\
&& \!\!\!\! - \sum_{j=\frac{1}{2}}^{n-\frac{1}{2}} S_{j}^{-} f_{n-j}^{+}
- \sum_{i=0}^{N-1} \sum_{j=\frac{1}{2}}^{n-\frac{1}{2}}
{(p_{j}^{-,1^{-},i} L_{1} +
p_{j}^{-,2^{-},i} G_{\frac{1}{2}}^{+} G_{\frac{1}{2}}^{-}) L_{1}^{i}}
f_{n-j+i+1}^{+}  \nn \\
&& - \sum_{i=0}^{N-1}
\sum_{j=\frac{3}{2}}^{n+\frac{1}{2}}
( p_{j-\frac{1}{2}}^{-,3^{--},i} G_{\frac{1}{2}}^{+} +
p_{j-\frac{1}{2}}^{-,4,i} G_{\frac{1}{2}}^{-} )
L_{1}^{i}
f_{n-j+i+1}^{+} \nn
\\
&& \!\!\!\! + \sum_{j=\frac{1}{2}}^{n-\frac{1}{2}} S_{j}^{+} f_{n-j}^{-}
+ \sum_{i=0}^{N-1} \sum_{j=\frac{1}{2}}^{n-\frac{1}{2}}
{(p_{j}^{+,1^{+},i} L_{1} +
p_{j}^{+,2^{+},i} G_{\frac{1}{2}}^{+} G_{\frac{1}{2}}^{-}) L_{1}^{i}}
f_{n-j+i+1}^{-} \nn \\
&& + \sum_{i=0}^{N-1} \sum_{j=\frac{3}{2}}^{n+\frac{1}{2}}
(p_{j-\frac{1}{2}}^{+,3,i} G_{\frac{1}{2}}^{+} +
p_{j-\frac{1}{2}}^{+,4^{++},i} G_{\frac{1}{2}}^{-} )
L_{1}^{i}
f_{n-j+i+1}^{-}
\eea
\bea
Y^{-}_{r}\!\!\!\! &=& \!\!\! \sum_{m=1}^{r-\frac{1}{2}} S_{m} f_{r-m}^{-} +
\sum_{i=0}^{N-1} \sum_{m=1}^{r+\frac{1}{2}} \Pki f_{r-m+i+1}^{-}
\nn \\
&& \!\!\! + \sum_{j=\frac{1}{2}}^{r} S_{j}^{-} f_{r-j} +
\sum_{i=0}^{N-1}
\sum_{j=\frac{1}{2}}^{r} {(p_{j}^{-,1^{-},i} L_{1} +
p_{j}^{-,2^{-},i} G_{\frac{1}{2}}^{+} G_{\frac{1}{2}}^{-} )
L_{1}^{i}} f_{r-j+i+1} \nn \\
&& + \sum_{i=0}^{N-1} \sum_{j=\frac{3}{2}}^{r} (
p_{j-\frac{1}{2}}^{-,3^{--},i} G_{\frac{1}{2}}^{+} +
p_{j-\frac{1}{2}}^{-,4,i} G_{\frac{1}{2}}^{-} )
L_{1}^{i} f_{r-j+i+1} \nn \\
Y^{+}_{r}\!\!\!\! &=&\!\!\! \sum_{m=1}^{r-\frac{1}{2}} S_{m} f_{r-m}^{+} +
\sum_{i=0}^{N-1} \sum_{m=1}^{r+\frac{1}{2}} \Pki f_{r-m+i+1}^{+}
\nn \\
&& \!\!\! + \sum_{j=\frac{1}{2}}^{r} S_{j}^{+} f_{r-j} +
\sum_{i=0}^{N-1}
\sum_{j=\frac{1}{2}}^{r} {(p_{j}^{+,1^{+},i} L_{1} +
p_{j}^{+,2^{+},i} G_{\frac{1}{2}}^{+} G_{\frac{1}{2}}^{-} )
L_{1}^{i}} f_{r-j+i+1} \nn \\
&& + \sum_{i=0}^{N-1} \sum_{j=\frac{3}{2}}^{r}
( p_{j-\frac{1}{2}}^{+,3,i} G_{\frac{1}{2}}^{+} +
p_{j-\frac{1}{2}}^{+,4^{++},i} G_{\frac{1}{2}}^{-} )
L_{1}^{i} f_{r-j+i+1}
\eea
The initial conditions for the recursion can be found easily by
considering that we have no
singular vectors at level $0$. Hence, in order to decouple the singular
vector ${\cal N}$ we necessarily need $j_{0}$ and $\overline{j}_{0}$
to vanish. The only terms contributing to them are
$f_{0}$ and $\overline{f}_{0}$,
which are both multiples of the highest weight
vector: $f_{0}=c_{f} \ket{h_{j},q_{j}}$, $\overline{f}_{0}=c_{
\overline{f}} \ket{h_{j},q_{j}}$. Using equations (\ref{eq:j_f}) we find:
\bea
\left( \begin{array}{c} 0 \\ 0 \end{array} \right)
&=& {\cal S}_{h_{j},q_{j}} \left( \begin{array}{c} c_{f} \\ c_{\overline{f}}
\end{array} \right) , \label{eq:init}
\eea
where the matrix ${\cal S}_{h_{j},q_{j}}$ is taken at level $0$:
$L_{0}=h_{j}$ and $T_{0}=q_{j}$. According to (\ref{eq:init}), $( c_{f} ,
c_{\overline{f}})^{T}$ has to be an eigenvector of ${\cal S}_{h_{j},q_{j}}$
with respect to the eigenvalue $0$. This determines restrictions
on the allowed fusions to
be the ones where ${\cal S}_{h_{j},q_{j}}$ has an eigenvalue $0$.

If the recursion breaks down at a certain level $r$ because
$s^{+}_{h_{j}+r,q_{j}+1}$ or
$s^{-}_{h_{j}+r,q_{j}-1}$ happens
 to be $0$ then we can use (\ref{eq:j_f}) to compute
$j_{r}^{+}$ or $j_{r}^{-}$ respectively which will be singular. If at
a certain level $m$, ${\cal S}_{h_{j}+m,q_{j}}$ happens to be singular, we
get a singular vector by
using (\ref{eq:j_f}) to compute the combination of $j_{m}$ and
$\overline{j}_{m}$ which corresponds to the eigenvalue $0$ and hence
does not involve the unknown terms $f_{m}$ and $\overline{f}_{m}$.
This completes the recursion procedure. In the following section we
will outline an induction proof for it.

Although the recursion formulae look rather complicated, in the actual
calculation they turn out to simplify rather drastically.

\subsection{Proof of the recursion procedure}
We shall now forget about the heuristic arguments used in the section
above and summarise the recursion procedure: we use as initial
conditions $f_{0}$ and $\overline{f}_{0}$ according to equation
(\ref{eq:init}); we perform odd and even recursion steps for
$f_{n},\overline{f}_{n},f_{r}^{\pm}$ alternately according to
equations (\ref{eq:step_e}, \ref{eq:step_o}) until the procedure
breaks down at level $m$
because ${\cal S}_{h_{j}+m,q_{j}}, s^{+}_{h_{j}+m,q_{j}+1}$
or $s^{-}_{h_{j}+m,q_{j}-1}$ is not invertible any
more. The coefficients $f_{n},\overline{f}_{n},f_{r}^{\pm}$ $(n,r<m)$
obtained in this way satisfy the descent equations (\ref{eq:descent}),
the $j_{n},\overline{j}_{n},j_{r}^{\pm}$ $(n,r<m)$ vanish
and respectively $j_{m}^{\pm}$ or the $j_{m}, \overline{j}_{m}$
combination corresponding to the ${\cal S}_{h_{j}+m,q_{j}}$
eigenvalue $0$ at level
$m$ is a singular vector or vanishes completely.

We shall prove this by induction on the level $n$. For $n=0$ we have
chosen $j_{0}=\overline{j}_{0}$ equals $0$ by construction of $f_{0}$ and
$\overline{f}_{0}$ which are multiples of the highest weight vector
$\ket{h_{j},q_{j}}$ and hence satisfy the descent equations
(\ref{eq:descent}). As induction hypothesis we shall assume now:
\bea
& \det({\cal S}_{h_{j},q_{j}}) = 0 \nn \\
& \det({\cal S}_{h_{j}+k,q_{j}}) \neq 0 \;\;\; ,\; 0<k<n  \\
& s^{\pm}_{h_{j}+r,q_{j}\pm 1} \neq 0 \;\;\; , \; 0<r<n \nn
\eea
and the descent equations are satisfied by $f_{k},f_{r}^{\pm}$ and
$\overline{f}_{k}$ for $0 \leq k <n$ , $0<r<n$.

Assume $n$ is
integral\footnote{Similar arguments can be used if $n$ is half
integral.}. We show that either $\det({\cal S}_{h_{j}+n,q_{j}}) \neq 0$
which will imply the descent equations for $f_{n}$ and
$\overline{f}_{n}$, or $\det({\cal S}_{h_{j}+n,q_{j}})=0$ which will
imply the construction of a singular vector at level $n$.

As a first step we want to translate $G_{\frac{1}{2}}^{\pm}$ and
$L_{1}$ back to superderivative language. If we use the notation
\bea
\gamma_{1} &=& 2h_{1}z+z\theta^{+} \frac{\partial}{\partial
\theta^{+}} + z \theta^{-} \frac{\partial}{\partial \theta^{-}} +z^{2}
\partial_{z} +q_{1} \theta^{+}\theta^{-} \nn \\
\gamma_{\pm} &=& 2h_{1} \theta^{\pm} - z \frac{\partial}{\partial
\theta^{\mp}} + \theta^{\pm} z \partial_{z} \pm \theta^{+}\theta^{-}
\frac{\partial}{\partial \theta^{\mp}} \pm q_{1} \theta^{\pm} ,
\label{eq:gammas}
\eea
we find:
\bea
L_{1}^{i+1} \Phi_{h_{1},q_{1}}(Z) \ket{h_{2},q_{2}} &=& \gamma_{1}^{i+1}
\Phi_{h_{1},q_{1}}(Z) \ket{h_{2},q_{2}} \nn \\
G_{\frac{1}{2}}^{\pm} L_{1}^{i}  \Phi_{h_{1},q_{1}}(Z) \ket{h_{2},q_{2}} &=&
\gamma_{1}^{i} \gamma_{\pm} \Phi_{h_{1},q_{1}}(Z) \ket{h_{2},q_{2}} \\
G_{\frac{1}{2}}^{+} G_{\frac{1}{2}}^{-} L_{1}^{i}
\Phi_{h_{1},q_{1}}(Z) \ket{h_{2},q_{2}} &=&
\gamma_{1}^{i} \gamma_{-} \gamma_{+} \Phi_{h_{1},q_{1}}(Z)
\ket{h_{2},q_{2}} . \nn
\eea
The left ideal which annihilates the highest weight state
$\ket{h_{2},q_{2}}$ is equally well spanned by
$\{(l_{m}+h_{2}z\delta_{m,1}), (t_{m}+q_{2}z\delta_{m,1}), g_{r}^{\pm},
G_{\frac{1}{2}}^{\pm}, (L_{0}-h_{2}), (T_{0}-q_{2});
m\in \bbbn, r \in \bbbn+\frac{1}{2}\}$,
using (\ref{eq:littleg}). Since ${\cal N}\ket{h_{2},q_{2}}$ is
singular, the above left ideal annihilates this vector. Hence,
and by considering the
commutation relations we get:
\bea
G_{\frac{1}{2}}^{+} {\cal N} &=& {\cal N} G_{\frac{1}{2}}^{+} + {\cal
N}^{1,+} (L_{0}-h_{2}) + {\cal N}^{2,+} (T_{0}-q_{2})
\nn \\
G_{\frac{1}{2}}^{-} {\cal N} &=& {\cal N} G_{\frac{1}{2}}^{-} + {\cal
N}^{1,-} (L_{0}-h_{2}) + {\cal N}^{2,-} (T_{0}-q_{2})
\label{eq:n_prime} \\
(t_{1}+q_{2}z) {\cal N} &=& {\cal N} (t_{1}+q_{2}z) +{\cal N}'^{-}
G_{\frac{1}{2}}^{+} +{\cal N}'^{+}  G_{\frac{1}{2}}^{-} + {\cal
N}'^{1} (L_{0}-h_{2}) + {\cal N}'^{2} (T_{0}-q_{2}) .\nn
\eea
It is easy to see that $\sigma(L_{0}-h_{2}) {\cal F}(Z)=
\sigma(T_{0}-q_{2}) {\cal F}(Z)=0$.
We apply the representation $\sigma$ (\ref{eq:sigma}) to
(\ref{eq:n_prime}). For simplicity we shall only be interested in the
contributions involving no $\theta^{\pm}$ terms, however, the following
considerations can equally well be done for them. We take the
expressions (\ref{eq:sigma},\ref{eq:sz}) for $\sigma(G_{\frac{1}{2}}^{+})
\sigma({\cal N}) {\cal F}(Z)$ and use equations (\ref{eq:gammas}) to
replace algebra generators by superderivatives. By considering the
commutation relations we interchange $\sigma(G_{\frac{1}{2}}^{+})$ and
$\sigma({\cal N})$ and obtain a form according to
(\ref{eq:n_prime}):\footnote{$S_{0;L_{0}+\frac{1}{2},T_{0}-1}$ denotes
$S_{0}$ in which we have replaced $L_{0}$ by $L_{0}+\frac{1}{2}$ and
$T_{0}$ by $T_{0}-1$ etc.}
\bea
\sigma(G_{\frac{1}{2}}^{+}) \sigma({\cal N}) {\cal F}(Z) &=&
\Bigl\{ (G_{\frac{1}{2}}^{+}+z\frac{\partial}{\partial \theta^{-}})
 (\frac{1}{z^{N}} S_{0} + \sum_{i=0}^{N-1}
\frac{1}{z^{N+i+1}} (p_{0}^{1,i} L_{1}+p_{0}^{2,i} G_{\frac{1}{2}}^{+}
G_{\frac{1}{2}}^{-})L_{1}^{i} + \ldots) \nn \\
&& + {\cal O}(\theta^{\pm}) + {\cal
O}(\theta^{+}\theta^{-}) \Bigr\} {\cal F}(Z)  \nn \\
&& =\Bigl\{
( \frac{1}{z^{N}} S_{0;L_{0}+\frac{1}{2},T_{0}-1} + \sum_{i=0}^{N-1}
\frac{1}{z^{N+i+1}} (p_{0;L_{0}+\frac{1}{2},T_{0}-1}^{1,i}
\gamma_{1}^{i+1} \nn \\
&& + p_{0;L_{0}+\frac{1}{2},T_{0}-1}^{2,i}
\gamma_{1}^{i} \gamma_{-} \gamma_{+})+\ldots)
(G_{\frac{1}{2}}^{+}+z\frac{\partial}{\partial \theta^{-}})
\nn \\
&& + {\cal
O}(\theta^{\pm})  + {\cal
O}(\theta^{+}\theta^{-}) \Bigr\} {\cal F}(Z) .
\eea
Hence, using the induction hypothesis
$G_{\frac{1}{2}}^{+}f_{k}+f_{k-\frac{1}{2}}^{+}=0$ for $k<n$:
\bea
\sigma(G_{\frac{1}{2}}^{+}) {\cal J}(Z) &=&
 \Bigl\{ \frac{1}{z^{N}} S_{0;L_{0}+\frac{1}{2},T_{0}-1} + \sum_{i=0}^{N-1}
\frac{1}{z^{N+i+1}} (p_{0;L_{0}+\frac{1}{2},T_{0}-1}^{1,i}
\gamma_{1}^{i+1} \nn \\
&& + p_{0;L_{0}+\frac{1}{2},T_{0}-1}^{2,i}
\gamma_{1}^{i} \gamma_{-} \gamma_{+})+\ldots \Bigr\}
\Bigl( \sum_{k \geq n} z^{h_{j}-h_{1}-h_{2}+k} (G_{\frac{1}{2}}^{+} f_{k} +
f_{k-\frac{1}{2}}^{+})\Bigr) \nn \\
&&  + {\cal O}(\theta^{\pm})  + {\cal
O}(\theta^{+}\theta^{-}). \nn \\
\eea
The action of $\gamma_{1}^{i+1}$ and $\gamma_{\pm}$ on $\sum_{k
 \geq n} z^{h_{j}-h_{1}-h_{2}+k} (G_{\frac{1}{2}}^{+} f_{k} +
f_{k-\frac{1}{2}}^{+})$ reproduces the coefficients of $p_{0}^{1,i}$
and $p_{0}^{2,i}$ in equations (\ref{eq:s_coef}).
Comparing coefficients and doing the same for the $\theta^{\pm}$ and
$\theta^{+}\theta^{-}$ components, we find:
\bea
\left( \begin{array}{c}
G_{\frac{1}{2}}^{+} j_{n} \\ G_{\frac{1}{2}}^{+} \overline{j}_{n}
\end{array} \right)
 &=&
{\cal S}_{h_{j}+n,q_{j}}
 \left( \begin{array}{c}
G_{\frac{1}{2}}^{+}f_{n} + f_{n-\frac{1}{2}}^{+} \\
G_{\frac{1}{2}}^{+}\overline{f}_{n} -(h_{1}+L_{0}-h_{2}+\frac{1}{2}
+q_{1}) f_{n-\frac{1}{2}}^{+}
\end{array} \right) \label{eq:pf1}
\eea
Suppose ${\rm det}{\cal S}$ at level $n$ is non trivial, we have $j_{n}$ and
$\overline{j}_{n}$ equal to $0$ due to the definition of $f_{n}$ and
$\overline{f}_{n}$. Hence, $G_{\frac{1}{2}}^{+}f_{n}
=-f_{n-\frac{1}{2}}^{+}$ and $G_{\frac{1}{2}}^{+}
\overline{f}_{n} =(h_{1}+L_{0}-h_{2}+\frac{1}{2}
+q_{1}) f_{n-\frac{1}{2}}^{+}$. However, if ${\rm det}{\cal
S}_{h_{j}+n,q_{j}}$
at level $n$
does vanish, we multiply (\ref{eq:pf1}) by the
inverse basis transformation ${\cal
T}$, which transforms ${\cal S}_{h_{j}+n,q_{j}}$ in its Jordan normal form:
\bea
{\cal T}^{-1}
\left( \begin{array}{c}
G_{\frac{1}{2}}^{+} j_{n} \\ G_{\frac{1}{2}}^{+} \overline{j}_{n}
\end{array} \right)
&=& \left( \begin{array}{cc}
0 & 0 \\
\:\! * & *
\end{array} \right) {\cal T}^{-1}
 \left( \begin{array}{c}
G_{\frac{1}{2}}^{+}f_{n} + f_{n-\frac{1}{2}}^{+} \\
G_{\frac{1}{2}}^{+}\overline{f}_{n} -(h_{1}+L_{0}-h_{2}+\frac{1}{2}
+q_{1}) f_{n-\frac{1}{2}}^{+}
\end{array} \right)
\eea
Hence, we find for the first component of
${\cal T}^{-1}\left( \begin{array}{c}
j_{n} \\ \overline{j}_{n} \end{array} \right)$, $\Psi$ say, that
$G_{\frac{1}{2}}^{+} \Psi =0$.

Similar considerations for $G_{\frac{1}{2}}^{-}$ and $t_{1}+q_{2}z$
lead to descent equations for $G_{\frac{1}{2}}^{-}$ and $T_{1}$, or
to $G_{\frac{1}{2}}^{+}\Psi=0$ and $T_{1}\Psi=0$ depending whether
${\cal S}_{h_{j}+n,q_{j}}$ is invertible or not. Since
$G_{\frac{1}{2}}^{+}, G_{\frac{1}{2}}^{-}$ and $T_{1}$ generate ${\cal
A}_{+}$ this is already sufficient.

The above considerations turn out to be even easier if $n$ is taken to be
positive, half integral, and they lead to the same result depending
whether $s^{+}_{h_{j}+n,q_{j}+1}$ or $s^{-}_{h_{j}+n,q_{j}-1}$ is
invertible or not. Whichever fails to be invertible leads to a
singular vector $j_{n}^{+}$ or $j_{n}^{-}$ respectively, exactly in
the same way as for $\Psi$.
This completes the proof\footnote{This does not
exclude that the recursively computed vector is identical to
$0$. However, the non triviality can usually be seen by looking at
the coefficients of the vector explicitly.}.

\section{Explicit formulae for selected singular vectors}

\subsection{Positive charged singular vectors}
We start off using the singular vector (\ref{eq:svec_12}) of the
representation generated by the field $\Phi_{h_{1,2},q'}$ and
investigate possible fusions with fields $\Phi_{h_{k},q}$
to give fields $\Phi_{h_{\tilde{k}},\tilde{q}}$. Factorising the
recursion determinant (\ref{eq:rec_s}) at level $0$,
we find as allowed fusions:
\bea
& \Phi_{h_{-k},-1-q} \otimes \Phi_{h_{1,2},1+2q} \rightarrow
\Phi_{h_{k},q} \label{eq:odd_p} \\
& \Phi_{h_{-k},1-q} \otimes \Phi_{h_{1,2},-1+2q} \rightarrow
\Phi_{h_{k},q} \label{eq:odd_m}
\eea
Using either of these two fusions and looking at the recursion
formulae shows very easily that if we take the fusion
(\ref{eq:odd_p}) the procedure breaks down for positive $k$,
 whilst for fusion
(\ref{eq:odd_m}) it breaks down for negative $k$. This allows us to give
explicit expressions for all charged singular vectors, which will
be done in this and the following section.

We first consider the fusion (\ref{eq:odd_p}), and put $k$ positive,
half integral. The
recursion formulae given in section \ref{sctn:rec} shows that the only
combinations of generators which do appear are
\bea
{\cal L}^{+}_{-1} = \frac{1}{2} \frac{t}{q+1} L_{-1} + \frac{1}{4}
\frac{t}{q(q+1)} G_{-\frac{1}{2}}^{+} G_{-\frac{1}{2}}^{-} + T_{-1} \: ,
\nn
\eea
as well as $G_{-\frac{1}{2}}^{+}$ and $G_{-\frac{1}{2}}^{-}$.

Since the recursion formulae involve in this case only terms which
differ in level at most by one, we can express the recursion procedure
as a product of four by four matrices. We only have to make sure to
choose the right initial conditions and finally to project out the
singular vector correctly from the four-component vector
$(f_{k-\frac{1}{2}},\overline{f}_{k-\frac{1}{2}},
f_{k-1}^{-},f_{k-1}^{+})^{T}$.

The ``even'' recursion step is given by
\bea
{\cal E}^{+}(n)=\left( \begin{array}{cccc}
e^{+}_{1,1} {\cal L}^{+}_{-1} &
e^{+}_{1,2} {\cal L}^{+}_{-1} &
e^{+}_{1,3} G_{-\frac{1}{2}}^{+} &
e^{+}_{1,4} G_{-\frac{1}{2}}^{-} \\
e^{+}_{2,1} {\cal L}^{+}_{-1} &
e^{+}_{2,2} {\cal L}^{+}_{-1} &
e^{+}_{2,3} G_{-\frac{1}{2}}^{+} &
e^{+}_{2,4} G_{-\frac{1}{2}}^{-} \\
0 & 0 & 1 & 0 \\
0 & 0 & 0 & 1
\end{array} \right)
& , & \left(
\begin{array}{c}
f_{n} \\ {\overline{f}_{n}} \\ f_{n-\frac{1}{2}}^{-} \\
f_{n-\frac{1}{2}}^{+}
\end{array} \right) = {\cal E}^{+}(n) \left(
\begin{array}{c}
f_{n-1} \\ \overline{f}_{n-1} \\ f_{n-\frac{1}{2}}^{-} \\
f_{n-\frac{1}{2}}^{+}
\end{array} \right) \: , \nn \\
\eea
where the matrix entries are
\bea
 e^{+}_{1,1}&=&\frac{t(2q+1)(2(k-n)+1)+4q(q+1)}{2nt(t(2k-n)+2(q+1))} \nn \\
 e^{+}_{1,2}&=&-\frac{1}{n(t(2k-n)+2(q+1))} \nn \\
 e^{+}_{1,3}&=&\frac{t(2(k-n)+1)}{4nq(t(2k-n)+2(q+1))} \nn \\
 e^{+}_{1,4}&=&-\frac{t(2(k-n)+1)+4(q+1)}{4n(q+1)(t(2k-n)+2(q+1))} \nn \\
 e^{+}_{2,1}&=&-\frac{4qt(q+1)(4k^{2}t+4n-t+4kq)+4t^{2}(k-n)^{2}
 +16q(q+1)^{2}(kt+q)-t^{2}}{4nt^{2}(t(2k-n)+2(q+1))} \nn \\
 e^{+}_{2,2}&=&\frac{t(2q+1)(2(k-n)-1)+4q(q+1)}{2nt(t(2k-n)+2(q+1))}
\nn \\
 e^{+}_{2,3}&=&-\frac{t(2q+1)(4k^{2}-1)+4q(q+1)(2(k+n)+1)-4nt(2k-n)
}{8nq(t(2k-n)+2(q+1))} \nn \\
 e^{+}_{2,4}&=&\!\! \frac{4nt^{2}(2k-n)+\! t^{2}(2q+1)(4k^{2}-1)+\! 8t(q+1)
(q(3k-n)+k+n)+\! 4(4q-t)(q+1)^{2}}{8nt(q+1)(t(2k-n)+2(q+1))\;\;\; .}  \nn
\eea

The ``odd'' recursion step can be found to be
\bea
{\cal T}^{+}(r)=\left( \begin{array}{cccc}
1 & 0 & 0 & 0 \\
0 & 1 & 0 & 0 \\
\frac{2q(q+1)-t(k-r)}{2(q+1)(t(k-r)+2q)} G_{-\frac{1}{2}}^{-} &
\frac{t}{2(q+1)(t(k-r)+2q)} G_{-\frac{1}{2}}^{-} &
-\frac{2q}{t(k-r)+2q} {\cal L}^{+}_{-1} & 0 \\
-\frac{2q(q+1)-t(k-r)}{2qt(k-r)} G_{-\frac{1}{2}}^{+} &
\frac{1}{2q(k-r)} G_{-\frac{1}{2}}^{+} &
0 & -\frac{2(q+1)}{t(k-r)} {\cal L}^{+}_{-1}
\end{array} \right)   \nn \\
\eea
\bea
& \left(
\begin{array}{c}
f_{r-\frac{1}{2}} \\ {\overline{f}_{r-\frac{1}{2}}} \\ f_{r}^{-} \\
f_{r}^{+}
\end{array} \right) = {\cal T}^{+}(r) \left(
\begin{array}{c}
f_{r-\frac{1}{2}} \\ \overline{f}_{r-\frac{1}{2}} \\ f_{r-1}^{-} \\
f_{r-1}^{+}
\end{array} \right) . \nn
\eea

Finally, we find as initial conditions
\bea
& \left(
\begin{array}{c}
f_{0} \\ {\overline{f}_{0}} \\ f_{-\frac{1}{2}}^{-} \\
f_{-\frac{1}{2}}^{+}
\end{array} \right) = \psi_{0}^{+} = \left(
\begin{array}{c}
-2t \ket{h_{k},q} \\ (4q(q+1)+(2q+1)(2k-1)t) \ket{h_{k},q}  \\ 0 \\ 0
\end{array} \right) ,
\eea
where $\ket{h_{k},q}$ denotes the highest weight vector,
and we take
\bea
&
\Psi_{k} = \underbrace{
\left( \frac{1}{2} G_{-\frac{1}{2}}^{+} , - \frac{1}{4}
\frac{t}{q(q+1)} G_{-\frac{1}{2}}^{+} , 0 , {\cal L}^{+}_{-1} \right)
}_{={\cal W}^{+}}
\left(
\begin{array}{c}
f_{k-\frac{1}{2}} \\ \overline{f}_{k-\frac{1}{2}} \\ f_{k-1}^{-} \\
f_{k-1}^{+}
\end{array} \right) .
\eea

Altogether, we can write the $+1$ charged singular vector at level $k$
in the Verma
module ${\cal V}_{h_{k},q}$ with $k$ in $\bbbn_{0}+\frac{1}{2}$
in the form
\bea
\Psi_{k} &=& {\cal W}^{+} {\cal E}^{+}(k-\frac{1}{2}) {\cal
T}^{+}(k-1)
{\cal E}^{+}(k-\frac{3}{2}) {\cal T}^{+}(k-2) \ldots {\cal E}^{+}(1)
{\cal T}^{+}(\frac{1}{2}) \psi_{0}^{+} .
\eea

\subsection{Negative charged singular vectors}
Using the fusion (\ref{eq:odd_m}) allows us to compute the singular
vectors of the Verma modules ${\cal V}_{h_{k},q}$ for $k$ negative, half
integral. Similarly to the previous section we find
\bea
{\cal L}^{-}_{-1} = \frac{t}{2q} L_{-1} + \frac{1}{4}
\frac{t}{q(q-1)} G_{-\frac{1}{2}}^{+} G_{-\frac{1}{2}}^{-} + T_{-1} \; ,
\nn
\eea
\bea
{\cal E}^{-}(n)=\left( \begin{array}{cccc}
e^{-}_{1,1} {\cal L}^{-}_{-1} &
e^{-}_{1,2} {\cal L}^{-}_{-1} &
e^{-}_{1,3} G_{-\frac{1}{2}}^{+} &
e^{-}_{1,4} G_{-\frac{1}{2}}^{-} \\
e^{-}_{2,1} {\cal L}^{-}_{-1} &
e^{-}_{2,2} {\cal L}^{-}_{-1} &
e^{-}_{2,3} G_{-\frac{1}{2}}^{+} &
e^{-}_{2,4} G_{-\frac{1}{2}}^{-} \\
0 & 0 & 1 & 0 \\
0 & 0 & 0 & 1
\end{array} \right)
& , & \left(
\begin{array}{c}
f_{n} \\ {\overline{f}_{n}} \\ f_{n-\frac{1}{2}}^{-} \\
f_{n-\frac{1}{2}}^{+}
\end{array} \right) = {\cal E}^{-}(n) \left(
\begin{array}{c}
f_{n-1} \\ \overline{f}_{n-1} \\ f_{n-\frac{1}{2}}^{-} \\
f_{n-\frac{1}{2}}^{+}
\end{array} \right) , \nn \\
\eea
\bea
 e^{-}_{1,1}&=&\frac{t(2q-1)(2(k+n)-1)+4q(q-1)}
{2nt(t(2k+n)+2(q-1))} \nn \\
 e^{-}_{1,2}&=&\frac{1}{n(t(2k+n)+2(q-1))} \nn \\
 e^{-}_{1,3}&=&\frac{t(2(k+n)-1)+4(q-1)}{4n(q-1)(t(2k+n)+2(q-1))} \nn \\
 e^{-}_{1,4}&=&-\frac{t(2(k+n)-1)}{4nq(t(2k+n)+2(q-1))} \nn \\
 e^{-}_{2,1}&=&\frac{4qt(q-1)(4k^{2}t+4n-t+4kq)+4t^{2}(k+n)^{2}+16q(q-1)^{2}
              (kt+q)-t^{2}}{4nt^{2}(t(2k+n)+2(q-1))} \nn \\
 e^{-}_{2,2}&=&\frac{t(2q-1)(2(k+n)+1)+4q(q-1)}
{2nt(t(2k+n)+2(q-1))} \nn \\
 e^{-}_{2,3}&=&\!\! \frac{4nt^{2}(2k+n)+\!
t^{2}(2q-1)(4k^{2}-1)
+\! 8t(q-1) (q(3k+n)-k+n)+\! 4(4q+t)(q-1)^{2}}{8nt(q-1)
(t(2k+n)+2(q-1))} \nn \\
 e^{-}_{2,4}&=&-\frac{(t(2q-1)(4k^{2}-1)+4q(q-1)(2(k-n)-1)-4nt(2k+n)
}{8nq(t(2k+n)+2(q-1))} \nn ,
\eea
\bea
{\cal T}^{-}(r)=\left( \begin{array}{cccc}
1 & 0 & 0 & 0 \\
0 & 1 & 0 & 0 \\
-\frac{2q(q-1)+t(k+r)}{2qt(k+r)} G_{-\frac{1}{2}}^{-} &
-\frac{1}{2q(k+r)} G_{-\frac{1}{2}}^{-} &
\frac{2(q-1)}{t(k+r)} {\cal L}^{-}_{-1} & 0 \\
\frac{2q(q-1)+t(k+r)}{2(q-1)(t(k+r)+2q)} G_{-\frac{1}{2}}^{+} &
-\frac{t}{2(q-1)(t(k+r)+2q)} G_{-\frac{1}{2}}^{+} &
0 & \frac{2q}{t(k+r)+2q} {\cal L}^{-}_{-1}
\end{array} \right)   \nn \\
\eea
\bea
& \left(
\begin{array}{c}
f_{0} \\ {\overline{f}_{0}} \\ f_{-\frac{1}{2}}^{-} \\
f_{-\frac{1}{2}}^{+}
\end{array} \right) = \psi^{-}_{0} = \left(
\begin{array}{c}
2t \ket{h_{k},q} \\ (4q(q-1)+(2q-1)(2k+1)t) \ket{h_{k},q}  \\ 0 \\ 0
\end{array} \right) ,
\eea
\bea
&
\Psi_{k} = \underbrace{
\left(- \frac{1}{2} G_{-\frac{1}{2}}^{-} , - \frac{1}{4}
\frac{t}{q(q-1)} G_{-\frac{1}{2}}^{-} ,  {\cal L}^{-}_{-1},0 \right)
}_{={\cal W}^{-}}
\left(
\begin{array}{c}
f_{k-\frac{1}{2}} \\ \overline{f}_{k-\frac{1}{2}} \\ f_{k-1}^{-} \\
f_{k-1}^{+}
\end{array} \right) .
\eea

And finally, we can write down the $-1$ charged singular vector in the Verma
module ${\cal V}_{h_{k},q}$ $(-k \in \bbbn_{0}+\frac{1}{2})$ at level
$|k|$:
\bea
\Psi_{k} &=& {\cal W}^{-} {\cal E}^{-}(|k|-\frac{1}{2})
{\cal T}^{-}(|k|-1)
{\cal E}^{-}(|k|-\frac{3}{2}) {\cal T}^{-}(|k|-2) \ldots {\cal E}^{-}(1)
{\cal T}^{-}(\frac{1}{2}) \psi^{-}_{0} .
\eea

\subsection{Uncharged $(r,2)$ singular vectors} \label{sctn:unch}
In this section we use the recursion in order to produce
uncharged singular vectors $\Psi_{r,s}$. Factorising the recursion
determinant at level $0$ gives the allowed fusions
\bea
& \Phi_{h_{r,s},q-\tilde{q}} \otimes \Phi_{h_{1,2},\tilde{q}} \rightarrow
\Phi_{h_{r,s-2},q} + \Phi_{h_{r,s+2},q} .  \label{eq:even_12}
\eea
Factorising the recursion matrix at the level where we are looking for
the singular vector tells us that the recursion will break down
for the cases we consider
if and only if $s=0$. However, this is not at all surprising since we
expect to find another freedom in solving the descent equations if the
Verma module corresponding to $\Phi_{h_{r,s},q-\tilde{q}}$ is
reducible. In such a case we would have to use information about the
singular vector of ${\cal V}_{h_{r,s},q-\tilde{q}}$ as well, as
suggested by Bauer et al. \cite{bfiz2} for the Virasoro case.
Putting $s=0$ allows us to compute the singular vectors $\Psi_{r,2}$
in exactly the same manner as in the two previous sections via the
fusion\footnote{We still have the freedom to choose $\tilde{q}$
conveniently; we suggest $\tilde{q}=0$.}
\bea
& \Phi_{h_{r,0},q} \otimes \Phi_{h_{1,2},0} \rightarrow
  \Phi_{h_{r,2},q} + \ldots \;\;\; . \label{eq:fus_un}
\eea

We obtain for $\Psi_{r,2}$, $r \in \bbbn$:
\bea
{\cal L}_{-1}=t L_{-1} - t
G_{-\frac{1}{2}}^{+} G_{-\frac{1}{2}}^{-} + T_{-1} & , &
\psi_{0}= \left(
\begin{array}{c}
-t \ket{h_{r,2},q} \\ q \ket{h_{r,2},q} \\ 0 \\ 0
\end{array} \right) \; ,
\eea
\vbox{
${\cal E}(n) =$
\bea
\left( \!\!
\begin{array}{cccc}
-\frac{q}{n t^{2} (n-r)} {\cal L}_{-1} &
\frac{1}{nt(n-r)} {\cal L}_{-1} &
-\frac{t(2n-r-1)-2q}{2nt(n-r)} G_{-\frac{1}{2}}^{+} &
-\frac{t(2n-r-1)+2q}{2nt(n-r)} G_{-\frac{1}{2}}^{-} \\
\!\! \frac{nt^{2}(n-r)+q^{2}}{nt^{3}(n-r)} {\cal L}_{-1} &
-\frac{q}{nt^{2}(n-r)} {\cal L}_{-1} &
\left( \frac{qt(2n-r-1)-2q^{2}}{2nt^{2}(n-r)}-1 \right) G_{-\frac{1}{2}}^{+} &
\left( \frac{qt(2n-r-1)+2q^{2}}{2nt^{2}(n-r)}+1 \right) G_{-\frac{1}{2}}^{-} \\
0 & 0 & 1 & 0 \\
0 & 0 & 0 & 1
\end{array} \!\!
\right) \nn \\
\eea
}
\bea
{\cal T}(s)= \!\! \left(
\begin{array}{cccc}
1 & 0 & 0 & \!\!\!\!\!\!\!\!\!\!\! 0 \\
0 & 1 & 0 & \!\!\!\!\!\!\!\!\!\!\! 0 \\
\frac{-t(r-2s)}{2(q-1)+t(r-2s)} G_{-\frac{1}{2}}^{-} &
\frac{2t}{2(q-1)+t(r-2s)} G_{-\frac{1}{2}}^{-} &
\frac{2}{2(q-1)+t(r-2s)} {\cal L}_{-1} & \!\!\!\!\!\!\!\!\!\!\! 0 \\
\frac{t(r-2s)}{2(q+1)-t(r-2s)} G_{-\frac{1}{2}}^{+} &
\frac{2t}{2(q+1)-t(r-2s)} G_{-\frac{1}{2}}^{+} & 0 &
\!\!\!\!\!\!\!\!\!\! \frac{2}{2(q+1)-t(r-2s)} {\cal L}_{-1}
\end{array} \right) \nn \\
\eea
\bea
{\cal W}(n) &=& \left(- \frac{q}{t} {\cal L}_{-1} , {\cal L}_{-1} ,
-(\frac{t}{2}(n-1)-q) G_{-\frac{1}{2}}^{+} , -(\frac{t}{2}(n-1)+q)
G_{-\frac{1}{2}}^{-} \right)
\eea
\bea
\Psi_{r,2} &=& {\cal W}(r) {\cal T}(r-\frac{1}{2}) {\cal E}(r-1)
{\cal T}(r-\frac{3}{2}) {\cal E}(r-2) \ldots {\cal E}(1) {\cal
T}(\frac{1}{2}) \psi_{0} .
\eea

\subsection{General uncharged singular vectors}
Based on associativity arguments for the fusions found in the
previous sections we conjecture that the fusion
\bea
\Phi_{h_{0,s+2},q-\tilde{q}} \otimes \Phi_{h_{r+1,2},\tilde{q}} \rightarrow
\Phi_{h_{r,s},q} + \ldots
\eea
will give us an expression for the uncharged singular vector
$\Psi_{r,s}$, where we have to use the knowledge of the singular vector
$\Psi_{r,2}$ obtained in section \ref{sctn:unch}.
Another possible fusion we could take is
\bea
\Phi_{h_{r,0},q-\tilde{q}} \otimes \Phi_{h_{1,s},\tilde{q}} \rightarrow
\Phi_{h_{r,s},q} + \ldots \; ,
\eea
where $\Psi_{1,s}$ can be found using the
fusion $\Phi_{h_{0,s+2},q} \otimes \Phi_{h_{2,2},0} \rightarrow
\Phi_{h_{1,s},q} + \ldots\:$. However, $\Psi_{2,2}$ is at level two,
which makes the recursion procedure much more complicated.

\section{Changing the fusion point}

The fusion we were using in the last sections corresponds to the three
point function taken at the super point $Z=(z,\theta^{+},\theta^{-})$
and the origin:
\bea
& \bra{0} \Phi_{h_{j},q_{j}}(0) \Phi_{h_{1},q_{1}}(Z)
\Phi_{h_{2},q_{2}}(0) \ket{0} \;\; .
\eea
This choice was made for simplicity, however we could have chosen a
three point function which is based on three different points:
\bea
& \bra{0} \Phi_{h_{j},q_{j}}(Z_{f}) \Phi_{h_{1},q_{1}}(Z_{1})
\Phi_{h_{2},q_{2}}(Z_{2}) \ket{0} \;\; . \label{eq:3pf}
\eea
In fact choosing the mid point of $Z_{1}$ and $Z_{2}$ as fusion point
$Z_{f}$ was a crucial
step in the Virasoro case to calculate
the precise formulae of Benoit and Saint-Aubin \cite{bsa1}. The same
trick also works for the affine algebra \cite{bauer} $A^{(1)}_1$.
Although the singular vectors are independent of the
chosen fusion point, the recursion matrices depend on
it. In this section we investigate this dependence and calculate it
explicitly for the vectors $\Psi_{r,2}$ choosing the mid point as
fusion point.

\subsection{Operator product expansion}
Exactly as before we need to find the action of the algebra generators
on the coefficients of the operator product expansion, the so called descent
equations. For simplicity,
we will only choose fusion points $Z_{f}$ on the line connecting
the two super points $Z_{1}$ and $Z_{2}$. This simplifies the operator
product expansion so that it only depends on the
super co-ordinates $Z_{12}$ and $\theta_{12}^{\pm}$:
\bea
\Phi_{h_{1},q_{1}}(Z_{1}) \: \Phi_{h_{2},q_{2}}(Z_{2}) \ket{0} &=&
 \sum_{j \in {\cal J} } C_{12}^{j}
Z_{12}^{h_{j}-h_{1}-h_{2}} \ket{\psi_{j}(Z_{12},
\theta_{12}^{\pm},Z_{f})} \;\; ,\nn \\
\ket{\psi_{j}(Z_{12},\theta_{12}^{\pm},Z_{f})} &=&
\sum_{n \in \bbbn_{0}} Z_{12}^{n} \ket{h_{j}+n,q_{j}} +
\theta_{12}^{+} \sum_{r\in \bbbn_{0} +\frac{1}{2}} Z_{12}^{r-\frac{1}{2}}
\ket{h_{j}+r,q_{j}-1}  \\
&& + \theta_{12}^{-} \sum_{r\in \bbbn_{0} +\frac{1}{2}} Z_{12}^{r-\frac{1}{2}}
\ket{h_{j}+r,q_{j}+1} +
\theta_{12}^{+} \theta_{12}^{-} \sum_{n\in \bbbn_{0}} Z_{12}^{n-1}
\overline{\ket{h_{j}+n,q_{j}}} . \nn
\eea
It is worth remarking that the vectors $\ket{h_{j}+n,q_{j}},
\ket{h_{j}+r,q_{j}\pm 1}$ and $\overline{\ket{h_{j}+n,q_{j}}}$ still
depend on the fusion point.
Choosing a different fusion point is the result of conjugating the
fields by the generators $L_{-1}$ and $G_{-\frac{1}{2}}^{\pm}$:
\bea
\Phi(Z) &=& e^{zL_{-1}} e^{-\theta^{+}G^{-}_{-1/2}}
e^{-\theta^{-}G^{+}_{-1/2}} \Phi(0)
e^{\theta^{-}G^{+}_{-1/2}} e^{\theta^{+}G^{-}_{-1/2}}
e^{-zL_{-1}}
\eea
The conjugated commutation relations
\bea
\ [L_{m}(Z_{0}),\Phi (Z_{2})] &=&
\Bigl\{ h(m+1) Z_{20}^{m} + \frac{1}{2} (m+1) Z_{20}^{m}
(\theta_{20}^{+} D_{2}^{-} + \theta_{20}^{-} D_{2}^{+}) +Z_{20}^{m+1}
\partial_{z_{2}} \nn \\
&& +\frac{q}{2} \theta_{20}^{+} \theta_{20}^{-} Z_{20}^{m-1} m (m+1) \Bigr\}
\Phi(Z_{2}) \nn \\
\ [G_{r}^{\pm}(Z_{0}),\Phi (Z_{2})]
&=& \Bigl\{ 2 h (r+\frac{1}{2}) \theta_{20}^{\pm}
Z_{20}^{r-\frac{1}{2}} -Z_{20}^{r+\frac{1}{2}}
D_{2}^{\pm} \pm \theta_{20}^{+} \theta_{20}^{-}
(r+\frac{1}{2}) Z_{20}^{r-\frac{1}{2}} D_{2}^{\pm} \nn \\
&& + 2 \theta_{20}^{\pm} Z_{20}^{r+\frac{1}{2}} \partial_{z_{2}} \pm q
\theta_{20}^{\pm} Z_{20}^{r-\frac{1}{2}} (r+\frac{1}{2}) \Bigr\}
\Phi(Z_{2}) \label{eq:phi_cr_eta} \\
\ [T_{m}(Z_{0}),\Phi(Z_{2})]
&=& \Bigl\{ 2 h \theta_{20}^{+} \theta_{20}^{-} m Z_{20}^{m-1} +
Z_{20}^{m} (\theta_{20}^{-} D_{2}^{+} - \theta_{20}^{+} D_{2}^{-}) +
2 \theta_{20}^{+} \theta_{20}^{-}
Z_{20}^{m} \partial_{z_{2}} \nn \\
&& + q Z_{20}^{m} \Bigr\} \Phi(Z_{2}) , \nn
\eea
lead us to the conjugated version of the representation $\sigma$:
\bea
\sigma(L_{m}(Z_{0})) \Phi(Z_{2}) &=& \Bigl\{ L_{m}(Z_{0})-h(m+1) Z_{20}^{m}
-Z_{20}^{m+1} \partial_{z_{2}} -
\frac{1}{2} (m+1) Z_{20}^{m} (\theta_{20}^{+} D_{2}^{-}
 +\theta_{20}^{-} D_{2}^{+}) \nn \\
&& -\frac{q}{2} m(m+1) \theta_{20}^{+} \theta_{20}^{-} Z_{20}^{m-1}
\Bigr\} \Phi(Z_{2}) \nn \\
\sigma(G_{r}^{\pm}(Z_{0})) \Phi(Z_{2}) &=& \Bigl\{ G_{r}^{\pm}(Z_{0})
 -2h(r+\frac{1}{2}) \theta_{20}^{\pm}
Z_{20}^{r-\frac{1}{2}} \mp (r+\frac{1}{2}) \theta_{20}^{+} \theta_{20}^{-}
Z_{20}^{r-\frac{1}{2}} D_{2}^{\pm} \label{eq:sigma_eta} \\
&& + Z_{20}^{r+\frac{1}{2}} D_{2}^{\pm} -2 \theta_{20}^{\pm}
Z_{20}^{r+\frac{1}{2}}
\partial_{z_{2}} \mp q (r+\frac{1}{2}) Z_{20}^{r-\frac{1}{2}}
\theta_{20}^{\pm} \Bigr\} \Phi(Z_{2}) \nn \\
\sigma(T_{m}(Z_{0})) \Phi(Z_{2}) &=& \Bigl\{ T_{m}(Z_{0})
 - 2hm \theta_{20}^{+} \theta_{20}^{-} Z_{20}^{m-1} -Z_{20}^{m}
(\theta_{20}^{-} D_{2}^{+} - \theta_{20}^{+} D_{2}^{-}) \nn \\
&& - 2 \theta_{20}^{+} \theta_{20}^{-} Z_{20}^{m}
\partial_{z_{2}} -q Z_{20}^{m} \Bigr\} \Phi(Z_{2}) \;\; . \nn
\eea

We classify the fusion point by a parameter $\eta$ which reflects
its position on the super line connecting
$Z_{1}=(z_{1},\theta^{+}_{1},\theta_{1}^{-})$ and
$Z_{2}=(z_{2},\theta^{+}_{2},\theta_{2}^{-})$:
\bea
Z_{f}=(z_{f},\theta^{+}_{f},\theta^{-}_{f}) &=&
(z_{2}+\eta (z_{1}-z_{2}), \theta_{2}^{+}+\eta
(\theta_{1}^{+}-\theta_{2}^{+}),
\theta_{2}^{-}+\eta
(\theta_{1}^{-}-\theta_{2}^{-})) \;\; .
\eea
A simple calculation shows how to relate the super co-ordinates:
\bea
Z_{1f} = (1-\eta) Z_{12} & \;\;\;\; & \theta_{1f}^{\pm} = (1-\eta)
\theta_{12}^{\pm} \nn \\
Z_{2f} = -\eta Z_{12} & \;\;\;\; & \theta_{2f}^{\pm} = -\eta
\theta_{12}^{\pm} \;\;\; .
\eea
Taking $\eta=0$ corresponds to choosing the point $Z_{2}$ as fusion point
which is exactly what we did in the previous sections, $\eta = \frac{1}{2}$ and
$\eta = 1$ correspond to the mid point and $Z_{1}$ respectively.
For the mid point the
coefficients $\ket{h_{j}+n,q_{j}},
\ket{h_{j}+r,q_{j}\pm 1}$ and $\overline{\ket{h_{j}+n,q_{j}}}$
only pick up sign factors according to their level by interchanging
$(h_{1},q_{1})$ and $(h_{2},q_{2})$, since $Z_{1f}=-Z_{2f}$
and $\theta_{1f}^{\pm}= - \theta_{2f}^{\pm}$.
This is why the mid point is distinguished.

\subsection{Descent equations}
The main difference to the descent equations derived above is that by
acting with the generators on the OPE we obtain contributions not only
from $\Phi_{1}$ but also from $\Phi_{2}$:
\bea
A(Z_{f}) \Phi_{1}(Z_{1}) \Phi_{2}(Z_{2}) &=&
 [A(Z_{f}),\Phi_{1}(Z_{1})] \Phi_{2}(Z_{2}) +
\Phi_{1}(Z_{1})  [A(Z_{f}),\Phi_{2}(Z_{2})] \;\; ,
\eea
where $A$ stands for an algebra generator.
For $\eta=0$ it is easy to see that the second part contributes
to $L_{0}$ and $T_{0}$ only.
The descent equations for $\eta \in
\bbbc$ are:

\bea \!\!
\begin{array}{lcl}
 L_{m} \: \ket{h_{j}+n,q_{j}} &=&
\Bigl\{ (m+1) (h_{1} (1-\eta)^{m} +h_{2} (-\eta)^{m})
+((1-\eta)^{m+1}-(-\eta)^{m+1}) \\[.8mm]
&& (h_{j}-h_{1}-h_{2}+n-m)\Bigr\}
\ket{h_{j}+n-m,q_{j}} \\[.8mm]
&& +(1-\eta) \eta ((1-\eta)^{m} - (-\eta)^{m}) L_{-1}
\ket{h_{j}+n-m-1,q_{j}} \\[.8mm]
 L_{m} \: \ket{h_{j}+r,q_{j}\pm 1} &=&
\Bigl\{ (m+1) (h_{1} (1-\eta)^{m} +h_{2} (-\eta)^{m})
+((1-\eta)^{m+1}-(-\eta)^{m+1}) \\[.8mm]
&& (h_{j}-h_{1}-h_{2}+r-\frac{m}{2})\Bigr\}
\ket{h_{j}+n-m,q_{j}\pm 1}
 \\[.8mm]
&& -(1-eta) \eta \frac{m+1}{2} ((1-\eta)^{m}-(-\eta)^{m})
G_{-\frac{1}{2}}^{\pm} \ket{h_{j}+n-m-\frac{1}{2},q_{j}} \\[.8mm]
&& +(1-\eta) \eta ((1-\eta)^{m} -(-\eta)^{m}) L_{-1}
\ket{h_{j}+n-m-1,q_{j}\pm 1}
\\[.8mm]
 L_{m} \: \overline{\ket{h_{j}+n,q_{j}}} &=&
\Bigl\{ (m+1) (h_{1} (1-\eta)^{m} +h_{2} (-\eta)^{m})
+((1-\eta)^{m+1}-(-\eta)^{m+1}) \\[.8mm]
&& (h_{j}-h_{1}-h_{2}+n)\Bigr\}
\overline{\ket{h_{j}+n-m,q_{j}}} \\[.8mm]
&& + \frac{m(m+1)}{2} (q_{1}(1-\eta)^{m+1}+q_{2}(-\eta)^{m+1})
\ket{h_{j}+n-m,q_{j}}\\[.8mm]
&& -\frac{m+1}{2} (1-\eta) \eta ((1-\eta)^{m}-(-\eta)^{m})
 G_{-\frac{1}{2}}^{+} \ket{h_{j}+n-m-\frac{1}{2},q_{j}-1} \\[.8mm]
&&  +\frac{m+1}{2} (1-\eta) \eta ((1-\eta)^{m}-(-\eta)^{m})
 G_{-\frac{1}{2}}^{-} \ket{h_{j}+n-m-\frac{1}{2},q_{j}+1}\\[.8mm]
&& (1-\eta) \eta ((1-\eta)^{m}-(-\eta)^{m}) L_{-1}
\overline{\ket{h_{j}+n-m-1,q_{j}}}
\\[.8mm]
G_{s}^{+} \: \ket{h_{j}+n,q_{j}} &=&
((-\eta)^{s+\frac{1}{2}}-(1-\eta)^{s+\frac{1}{2}})
 \ket{h_{j}+n-s,q_{j}+1} \\[.8mm]
&& +(1-\eta) \eta ((1-\eta)^{s-\frac{1}{2}} - (-\eta)^{s-\frac{1}{2}} )
G_{-\frac{1}{2}}^{+} \ket{h_{j}+n-s-\frac{1}{2},q_{j}} \\[.8mm]
G_{s}^{+} \: \ket{h_{j}+r,q_{j}-1} &=&
-\Bigl\{
2(s+\frac{1}{2})(h_{1}(1-\eta)^{s+\frac{1}{2}}+h_{2}(-\eta)^{s+\frac{1}{2}}
+(s+\frac{1}{2}) \\[.8mm]
&& (q_{1}(1-\eta)^{s+\frac{1}{2}}+q_{2}(-\eta)^{s+\frac{1}{2}})
+(1-2\eta)
((1-\eta)^{s+\frac{1}{2}}-(-\eta)^{s+\frac{1}{2}})  \\[.8mm]
&& (h_{j}-h_{1}-h_{2}+r-s)
\Bigr\} \ket{h_{j}+r-s,q_{j}} \\[.8mm]
&&  - ((1-\eta)^{s+\frac{1}{2}}-(-\eta)^{s+\frac{1}{2}})
\overline{\ket{h_{j}+r-s,q_{j}}} \\[.8mm]
&& - (1-\eta) \eta ((1-\eta)^{s+\frac{1}{2}}-(-\eta)^{s+\frac{1}{2}})
L_{-1} \ket{h_{j}+r-s-1,q_{j}} \\[.8mm]
&& + (1-\eta) \eta ((1-\eta)^{s-\frac{1}{2}}-(-\eta)^{s-\frac{1}{2}})
G_{-\frac{1}{2}}^{+} \ket{h_{j}+r-s-\frac{1}{2},q_{j}-1} \\[.8mm]
 G_{s}^{+} \: \ket{h_{j}+r,q_{j}+1} &=&
(1-\eta) \eta ((1-\eta)^{s-\frac{1}{2}} -(-\eta)^{s-\frac{1}{2}})
G_{-\frac{1}{2}}^{+} \ket{h_{j}+r-s-\frac{1}{2},q_{j}+1}  \\[.8mm]
 G_{s}^{+} \: \overline{\ket{h_{j}+n,q_{j}}} &=&
\Bigl\{2(s+\frac{1}{2}(h_{1}(1-\eta)^{s+\frac{1}{2}}+h_{2}
(-\eta)^{s+\frac{1}{2}})+(s+\frac{1}{2}) \\[.8mm]
&& (q_{1}(1-\eta)^{s+\frac{1}{2}}+q_{2}
(-\eta)^{s+\frac{1}{2}})
+(1-2\eta)((1-\eta)^{s+\frac{1}{2}}-(-\eta)^{s+\frac{1}{2}})  \\[.8mm]
&& (h_{j}-h_{1}-h_{2} +n-s-\frac{1}{2})
+(s+\frac{1}{2})((1-\eta)^{s+\frac{3}{2}} -(-\eta)^{s+\frac{3}{2}})
\Bigr\} \\[.8mm]
&& \ket{h_{j}+n-s,q_{j}+1} \\[.8mm]
&& +(1-\eta) \eta ((1-\eta)^{s+\frac{1}{2}}-(-\eta)^{s+\frac{1}{2}})
L_{-1} \ket{h_{j}+n-s-1,q_{j}+1} \\[.8mm]
&& - (s+\frac{1}{2}) (1-\eta) \eta
((1-\eta)^{s+\frac{1}{2}}-(-\eta)^{s+\frac{1}{2}})
G_{-\frac{1}{2}}^{+} \\[.8mm]
&& \ket{h_{j}+n-s-\frac{1}{2},q_{j}}  \\[.8mm]
&& + (1-\eta) \eta ((1-\eta)^{s-\frac{1}{2}} -(-\eta)^{s-\frac{1}{2}}
) G_{-\frac{1}{2}}^{+} \overline{\ket{h_{j}+n-s-\frac{1}{2},q_{j}}}
\\[.8mm]
\label{eq:descent_eta}
\end{array}
\eea
\bea \!\!
\begin{array}{lcl}
G_{s}^{-} \: \ket{h_{j}+n,q_{j}} &=&
((-\eta)^{s+\frac{1}{2}}-(1-\eta)^{s+\frac{1}{2}})
 \ket{h_{j}+n-s,q_{j}-1} \\[.8mm]
&& +(1-\eta) \eta ((1-\eta)^{s-\frac{1}{2}} - (-\eta)^{s-\frac{1}{2}} )
G_{-\frac{1}{2}}^{-} \ket{h_{j}+n-s-\frac{1}{2},q_{j}} \\[.8mm]
G_{s}^{-} \: \ket{h_{j}+r,q_{j}+1} &=&
-\Bigl\{
2(s+\frac{1}{2})(h_{1}(1-\eta)^{s+\frac{1}{2}}+h_{2}(-\eta)^{s+\frac{1}{2}}
-(s+\frac{1}{2}) \\[.8mm]
&& (q_{1}(1-\eta)^{s+\frac{1}{2}}+q_{2}(-\eta)^{s+\frac{1}{2}})
+(1-2\eta)
((1-\eta)^{s+\frac{1}{2}}-(-\eta)^{s+\frac{1}{2}})  \\[.8mm]
&& (h_{j}-h_{1}-h_{2}+r-s)
\Bigr\} \ket{h_{j}+r-s,q_{j}} \\[.8mm]
&&  + ((1-\eta)^{s+\frac{1}{2}}-(-\eta)^{s+\frac{1}{2}})
\overline{\ket{h_{j}+r-s,q_{j}}} \\[.8mm]
&& - (1-\eta) \eta ((1-\eta)^{s+\frac{1}{2}}-(-\eta)^{s+\frac{1}{2}})
L_{-1} \ket{h_{j}+r-s-1,q_{j}} \\[.8mm]
&& + (1-\eta) \eta ((1-\eta)^{s-\frac{1}{2}}-(-\eta)^{s-\frac{1}{2}})
G_{-\frac{1}{2}}^{-} \ket{h_{j}+r-s-\frac{1}{2},q_{j}+1} \\[.8mm]
 G_{s}^{-} \: \ket{h_{j}+r,q_{j}-1} &=&
(1-\eta) \eta ((1-\eta)^{s-\frac{1}{2}} -(-\eta)^{s-\frac{1}{2}})
G_{-\frac{1}{2}}^{-} \ket{h_{j}+r-s-\frac{1}{2},q_{j}-1}  \\[.8mm]
 G_{s}^{-} \: \overline{\ket{h_{j}+n,q_{j}}} &=&
\Bigl\{-2(s+\frac{1}{2}(h_{1}(1-\eta)^{s+\frac{1}{2}}+h_{2}
(-\eta)^{s+\frac{1}{2}})+(s+\frac{1}{2}) \\[.8mm]
&& (q_{1}(1-\eta)^{s+\frac{1}{2}}+q_{2}
(-\eta)^{s+\frac{1}{2}})
-(1-2\eta)((1-\eta)^{s+\frac{1}{2}}-(-\eta)^{s+\frac{1}{2}}) \\[.8mm]
&& (h_{j}-h_{1}-h_{2} +n-s-\frac{1}{2})
-(s+\frac{1}{2})((1-\eta)^{s+\frac{3}{2}} -(-\eta)^{s+\frac{3}{2}})
\Bigr\} \\[.8mm]
&& \ket{h_{j}+n-s,q_{j}-1} \\[.8mm]
&& -(1-\eta) \eta ((1-\eta)^{s+\frac{1}{2}}-(-\eta)^{s+\frac{1}{2}})
L_{-1} \ket{h_{j}+n-s-1,q_{j}-1} \\[.8mm]
&& + (s+\frac{1}{2}) (1-\eta) \eta
((1-\eta)^{s+\frac{1}{2}}-(-\eta)^{s+\frac{1}{2}})
G_{-\frac{1}{2}}^{-} \\[.8mm]
&& \ket{h_{j}+n-s-\frac{1}{2},q_{j}}  \\[.8mm]
&& + (1-\eta) \eta ((1-\eta)^{s-\frac{1}{2}} -(-\eta)^{s-\frac{1}{2}}
) G_{-\frac{1}{2}}^{-} \overline{\ket{h_{j}+n-s-\frac{1}{2},q_{j}}}
\\[.8mm]
 T_{m} \: \ket{h_{j}+n,q_{j}} &=&
(q_{1}(1-\eta)^{m}+q_{2}(-\eta)^{m})
 \ket{h_{j}+n-m,q_{j}} \\[.8mm]
 T_{m} \: \ket{h_{j}+r,q_{j}\pm 1} &=&
\Bigl\{ q_{1}(1-\eta)^{m}+q_{2}(-\eta)^{m}
\pm ((1-\eta)^{m+1}-(-\eta)^{m+1})) \Bigr\} \\[.8mm]
&& \ket{h_{j}+r-m,q_{j}\pm 1} \\[.8mm]
&& \mp (1-\eta) \eta ((1-\eta^{m}-(-\eta)^{m}) G_{-\frac{1}{2}}^{\pm}
\ket{h_{j}+r-m-\frac{1}{2},q_{j}\pm 1} \\[.8mm]
T_{m} \: \overline{\ket{h_{j}+n,q_{j}}} &=&
\Bigl\{ 2m(h_{1}(1-\eta)^{m+1} +h_{2}(-\eta)^{m+1}) -2 (1-\eta) \eta
 \\[.8mm]
&& ((1-\eta)^{m}-(-\eta)^{m}) (h_{j}-h_{1}-h_{2}+n-m) \Bigr\}
\ket{h_{j}+n-m,q_{j}} \\[.8mm]
&& + (q_{1}(1-\eta)^{m}+q_{2} (-\eta)^{m})
 \overline{\ket{h_{j}+n-m,q_{j}}}  \\[.8mm]
&& -(1-\eta) \eta ((1-\eta)^{m} -(-\eta)^{m}) G_{-\frac{1}{2}}^{+}
\ket{h_{j}+n-m-\frac{1}{2},q_{j}-1}  \\[.8mm]
&& -(1-\eta) \eta ((1-\eta)^{m} -(-\eta)^{m}) G_{-\frac{1}{2}}^{-}
\ket{h_{j}+n-m-\frac{1}{2},q_{j}+1} \;\; ,
\nn
\end{array}
\eea
where we have again put $q_{j}=q_{1}+q_{2}$, $m \in \bbbn$, $n \in
\bbbn_{0}$, $r,s \in \bbbn_{0}+\frac{1}{2}$ and all the generators are
taken at the fusion point.

\subsection{Expansion of the generators}
We are now able to describe the algorithm for this more general three
point function. We start off taking an allowed fusion which means we
take a non trivial three point function
\bea
\bra{0} \Phi_{h_{j},q_{j}} (Z_{f}) \Phi_{h_{1},q_{1}}(Z_{1})
\Phi_{h_{2},q_{2}}(Z_{2}) \ket{0} .
\eea
We insert as before a known operator mapping $\Phi_{h_{2},q_{2}} \ket{0}$ on a
singular vector:
\bea
\Psi &=& {\cal N}(Z_{2}) \Phi_{h_{2}q_{2}}(Z_{2}) \ket{0} .
\eea
Clearly all the generators in ${\cal N}$ have to be taken at the point
$Z_{2}$. We use the generalised representation
(\ref{eq:sigma_eta})
to obtain again:
\bea
0 & = & \bra{0} \Phi_{h_{j},q_{j}}(Z_{f}) \Phi_{h_{1},q_{1}}(Z_{1}) {\cal
N}(Z_{2}) \Phi_{h_{2},q_{2}}(Z_{2}) \ket{0} \nn \\
& = &  \bra{0} \Phi_{h_{j},q_{j}}(Z_{f}) \sigma ({\cal N}(Z_{2}))
\Phi_{h_{1},q_{1}}(Z_{1})
\Phi_{h_{2},q_{2}}(Z_{2}) \ket{0} .
\eea
We are facing now the problem that we want to act with the generators
in $\sigma ({\cal N}(Z_{2}))$ on the OPE of $\Phi_{h_{1},q_{1}}(Z_{1})
\Phi_{h_{2},q_{2}}(Z_{2})$. For this purpose it is necessary to write
the generators in $\sigma ({\cal N}(Z_{2}))$ which are taken at the
point $Z_{2}$ in terms of generators which are taken at the fusion
point $Z_{f}$.

Using the $N=2$ Taylor expansion
\bea
f(Z_{1}) =  \sum_{n=0}^{\infty} \frac{1}{n!} Z_{12}^{n}
\partial_{z_{2}}^{n} & \Bigl\{ &  1+\theta_{12}^{+} D_{2}^{-}
+\theta_{12}^{-} D_{2}^{+} +\theta_{12}^{+} \theta_{12}^{-} D_{2}^{+}
D_{2}^{-}
\nn \\
&& -\theta_{12}^{+} \theta_{12}^{-} \frac{\partial}{\partial
z_{2}} \Bigr\} f(Z_{2}) ,
\eea
we can expand the equations (\ref{eq:gen}) about the fusion point
which leads to:
\bea
L_{m}(Z_{2}) &=& \sum_{n=0}^{\infty} (\eta Z_{12})^{n} \Bigl\{ \binc{m+1}{n}
L_{m-n} (Z_{f}) -\eta \frac{m+1}{2} \theta_{12}^{+} \binc{m}{n}
G_{m-n-\frac{1}{2}}^{-}(Z_{f}) \nn \\
&& -\eta \frac{m+1}{2} \theta_{12}^{-} \binc{m}{n}
G_{m-n-\frac{1}{2}}^{+}(Z_{f}) \nn \\
&& + \eta^{2} \frac{m(m+1)}{2} \binc{m-1}{n}
\theta_{12}^{+} \theta_{12}^{-} T_{m-n-1}(Z_{f}) \Bigr\} \nn \\
G_{r}^{\pm}(Z_{f}) &=& \sum_{n=0}^{\infty} (\eta Z_{12})^{n} \Bigl\{
\binc{r+\frac{1}{2}}{n} G_{r-n}^{\pm}(Z_{f}) +2 \eta \theta_{12}^{\pm}
\binc{r+\frac{1}{2}}{n} L_{r-n-\frac{1}{2}}(Z_{f}) \nn \\
&& \pm \eta
\theta_{12}^{\pm} (r+\frac{1}{2}) \binc{r-\frac{1}{2}}{n}
T_{r-n-\frac{1}{2}}(Z_{f}) \label{eq:gen_exp} \\
&& \mp \eta^{2} \theta_{12}^{+} \theta_{12}^{-} (r+\frac{1}{2})
G_{r-n-1}^{\pm}(Z_{f}) \Bigr\} \nn \\
T_{m}(Z_{2}) &=& \sum_{n=0}^{\infty} (\eta Z_{12})^{n} \Bigl\{
\binc{m}{n} T_{m-n}(Z_{f}) +\eta \binc{m}{n} \theta_{12}^{+}
G_{m-n-\frac{1}{2}}(Z_{f}) \nn \\
&& -\eta \binc{m}{n} \theta_{12}^{-} G_{m-n-\frac{1}{2}}^{+}(Z_{f}) +2
\eta^{2} \binc{m}{n} \theta_{12}^{+} \theta_{12}^{-} L_{m-n-1}(Z_{f})
\;\; .
\Bigr\} \nn
\eea

We replace the generators in $\sigma ({\cal N}(Z_{2}))$ by the
generators taken at the fusion point $Z_{f}$ according to
(\ref{eq:gen_exp}) and then we follow the recursion procedure exactly
as in the previous sections. The only difference is that wherever we
find $z$ or $\theta^{\pm}$ in the expansions ${\cal F},{\cal J}$ and
${\cal S}$ we have to replace them by $Z_{12}$ or $\theta_{12}^{\pm}$
respectively. The proof follows by conjugation from the case where $\eta=0$.

\subsection{Example}
As mentioned earlier, we expect that the coefficients in ${\cal F}$
show certain symmetries under $(h_{1},q_{1})$ and $(h_{2},q_{2})$
exchange if we choose $\eta=\frac{1}{2}$. For this case we again would like
to give the result for the uncharged singular vectors
$\Psi_{r,2}$.

It turns out that the determinant of the recursion matrix ${\cal S}$
(\ref{eq:rec_s}) does
not depend on the parameter $\eta$. This reflects the fact that the fusion
rules are
independent of the chosen fusion point.
We again take the singular vector $\Psi_{1,2}$
 (\ref{eq:svec_12}) and use the fusion (\ref{eq:fus_un}) to perform the
recursion algorithm. In the Virasoro case we would have to deal at level one
only with operators $L_{-1}(Z_{2})$. Those expanded according to
(\ref{eq:gen_exp}) stay the same. We see that the generators
$G_{-\frac{1}{2}}^{\pm}(Z_{2})$ lead to a finite expansion whilst
$T_{-1}(Z_{2})$ involves an infinite series of $L_{-n}(Z_{f})$ and
$G_{-r}^{\pm}(Z_{f})$ generators. We will not be able to
write the vectors $\Psi_{r,2}$ in a closed expression other than a
product of matrices, as it was possible in the Virasoro case. However
the expressions appear more suitable to apply the
``analytic continuation method'' to obtain product expressions for all
singular vectors.

To write the result in a compact way, we introduce the four component
vectors:
\bea
F^{+}_{n} = \left( \begin{array}{c}
f_{n} \\ \overline{f}_{n} \\ f_{n+\frac{1}{2}}^{-} \\
f_{n+\frac{1}{2}}^{+} \end{array} \right) \;\;\;\;
&
F^{-}_{n} = \left( \begin{array}{c}
f_{n} \\ \overline{f}_{n} \\ f_{n-\frac{1}{2}}^{-} \\
f_{n-\frac{1}{2}}^{+} \end{array} \right) & \;\;\;\;
F^{-}_{0} = \left( \begin{array}{c}
-t \ket{h_{r,2},q} \\ q \ket{h_{r,2},q} \\ 0 \\ 0 \end{array} \right)
\;\; .
\eea
We define the matrices:
\bea
E_{1}(n) &=& \frac{1}{4(n-r)nt}
\left( \begin{array}{cccc}
e_{1}^{11}(n) & e_{1}^{12}(n) & e_{1}^{13}(n) & e_{1}^{14}(n) \\
e_{1}^{21}(n) & e_{1}^{22}(n) & e_{1}^{23}(n) & e_{1}^{24}(n) \\
0 & 0 & 4(n-r)nt & 0 \\
0 & 0 & 0 & 4(n-r)nt
\end{array} \right) \\
e_{1}^{11}(n) & = &
-((r+1)t+q+2-2nt)L_{-1}-4\frac{q}{t}T_{-1}+qG_{-\frac{1}{2}}^{+}
G_{-\frac{1}{2}}^{-} \nn \\
e_{1}^{12}(n) & = & 4T_{-1}-tG_{-\frac{1}{2}}^{+}
G_{-\frac{1}{2}}^{-}+tL_{-1} \nn \\
e_{1}^{13}(n) & = &
((r+1)t+2(q+1)-2nt)G_{-\frac{1}{2}}^{+} \nn \\
e_{1}^{14}(n) & = &
((r+1)t-2(q-1)-2nt)G_{-\frac{1}{2}}^{-} \nn \\
e_{1}^{21}(n) & = \!\! &
(-(tr+2q)n+(r+1)q+(q+2)\frac{q}{t}+n^{2}t)L_{-1}
+(n(n-r)+\frac{q^{2}}{t^{2}})
\nn \\ &&
(4T_{-1}-tG_{-\frac{1}{2}}^{+}G_{-\frac{1}{2}}^{-}) \nn \\
e_{1}^{22}(n) & = &
-4\frac{q}{t}T_{-1}+qG_{-\frac{1}{2}}^{+}G_{-\frac{1}{2}}^{-}
-qL_{-1} \nn \\
e_{1}^{23}(n) & = &
(2(tr+q)n-(r+1)q-2(q+1)\frac{q}{t}-2n^{2}t)
G_{-\frac{1}{2}}^{+} \nn \\
e_{1}^{24}(n) & = &
(-2(tr-q)n-(r+1)q+2(q-1)\frac{q}{t}+2n^{2}t)
G_{-\frac{1}{2}}^{-} \nn
\eea
\vbox{
$ E_{2}(n) = \frac{1}{2(n-r)nt}$
\bea
\left( \begin{array}{cccc}
L_{-2}-\frac{t}{8}L_{-1}^{2}+\frac{q}{t}T_{-2} &
-T_{-2} &
- G_{-\frac{3}{2}}^{+}+\frac{t}{4}L_{-1}G_{-\frac{1}{2}}^{+} &
- G_{-\frac{3}{2}}^{-}+\frac{t}{4}L_{-1}G_{-\frac{1}{2}}^{-} \\
-\frac{q}{t}L_{-2}+\frac{q}{8}L_{-1}^{2}-\frac{n(n-r)t^2+q^{2}}{t^{2}}T_{-2}
&
\frac{q}{t}T_{-2} &
\frac{q}{t}G_{-\frac{3}{2}}^{+}-\frac{q}{4}L_{-1}G_{-\frac{1}{2}}^{+} &
\frac{q}{t}G_{-\frac{3}{2}}^{-}-\frac{q}{4}L_{-1}G_{-\frac{1}{2}}^{-} \\
0 & 0 & 0 & 0 \\
0 & 0 & 0 & 0 \end{array} \right) \nn \\
\eea
}
\bea
E_{k}(n) \!\! &=& \!\! \frac{(-1)^{k}}{2^{k-1}(n-r)nt} \left(
\begin{array}{cccc}
 L_{-k} + \frac{q}{t}T_{-k} &
- T_{-k} &
- G_{-k+\frac{1}{2}}^{+} &
- G_{-k+\frac{1}{2}}^{-} \\
- \frac{q}{t}L_{-k}-\frac{n(n-r)t^{2}+q^{2}}{t^{2}}T_{-k} &
 \frac{q}{t}T_{-k} &
 \frac{q}{t} G_{-k+\frac{1}{2}}^{+}&
 \frac{q}{t} G_{-k+\frac{1}{2}}^{-} \\
0  & 0 & 0 & 0 \\
0  & 0 & 0 & 0
\end{array} \right)  \; {\rm , } \; k \geq 3 \nn \\
\eea
\bea
T_{1}(s) &=& \left( \begin{array}{cccc}
1 & 0 & 0 & 0 \\
0 & 1 & 0 & 0 \\
-\frac{(2st-tr-2)G_{-\frac{1}{2}}^{-}}{2(2ts-tr-2q+2)} &
-\frac{tG_{-\frac{1}{2}}^{-}}{2ts-tr-2q+2}  &
-\frac{4T_{-1}-tG_{-\frac{1}{2}}^{+}
G_{-\frac{1}{2}}^{-}}{2(2ts-tr-2q+2)} & 0 \\
-\frac{(2st-tr-2)G_{-\frac{1}{2}}^{+}}{2(2ts-tr+2q+2)} &
\frac{tG_{-\frac{1}{2}}^{+}}{2ts-tr+2q+2} &
0 & \frac{4T_{-1}-tG_{-\frac{1}{2}}^{+}
G_{-\frac{1}{2}}^{-}+2tL_{-1}}{2(2ts-tr+2q+2)} \end{array} \right) \nn \\
\\
T_{2}(s) &=& \left( \begin{array}{cccc}
0 & 0 & 0 & 0 \\
0 & 0 & 0 & 0 \\
\frac{tL_{-1}G_{-\frac{1}{2}}^{-}-4G_{-\frac{3}{2}}^{-}}{4(2ts-tr-2q+2)}
& 0 &  \frac{T_{-2}}{2ts-tr-2q+2} & 0 \\
\frac{tL_{-1}G_{-\frac{1}{2}}^{+}-4G_{-\frac{3}{2}}^{+}}{4(2ts-tr+2q+2)}
& 0 & 0 & - \frac{T_{-2}}{2ts-tr+2q+2} \end{array} \right)
\eea
\bea
T_{k}(s) &=& (-\frac{1}{2})^{k-2} \left(
\begin{array}{cccc}
0 & 0 & 0 & 0 \\
0 & 0 & 0 & 0 \\
- \frac{G_{-k+\frac{1}{2}}^{-}}{2ts-tr-2q+2} &
0 &
 \frac{T_{-k}}{2ts-tr-2q+2} & 0 \\
- \frac{G_{-k+\frac{1}{2}}^{+}}{2ts-tr+2q+2} &
0 & 0 &
- \frac{T_{-k}}{2st-rt+2q+2}
\end{array} \right) \;\; {\rm , } \; k \geq 3 \nn \\
\eea
\bea
W_{1}(r) & = & \left( \begin{array}{c}
-\frac{t}{2}((1-r)t+q+2)L_{-1}-2qT_{-1}+\frac{t}{2}qG_{-\frac{1}{2}}^{+}
G_{-\frac{1}{2}}^{-} \\
2tT_{-1}-\frac{t^{2}}{2}G_{-\frac{1}{2}}^{+}
G_{-\frac{1}{2}}^{-}+\frac{t^{2}}{2}L_{-1} \\
\frac{t}{2}((1-r)t+2(q+1)) G_{-\frac{1}{2}}^{+} \\
\frac{t}{2}((1-r)t-2(q-1)) G_{-\frac{1}{2}}^{-} \end{array}
\right) \\
W_{2}(r) & = & \left( \begin{array}{c}
-\frac{t^{2}}{8}L_{-1}^{2}+tL_{-2}+qT_{-2} \\
-tT_{-2} \\ -t
G_{-\frac{3}{2}}^{+}+\frac{t^{2}}{4}L_{-1}G_{-\frac{1}{2}}^{+}
\\ -t G_{-\frac{3}{2}}^{-} +\frac{t^{2}}{4}L_{-1}G_{-\frac{1}{2}}^{-}
\end{array} \right) \\
W_{k}(r) & = & (-\frac{1}{2})^{k-2} \left( \begin{array}{c}
tL_{-k}+qT_{-k} \\
- t T_{-k} \\
- tG_{-k+\frac{1}{2}}^{+} \\
- tG_{-k+\frac{1}{2}}^{-} \end{array} \right)
\;\;\;\; {\rm , } \; k \geq 3 \;\;\; .
\eea
These matrices enable us to write $F_{n}^{\pm}$ as:
\bea
F_{n}^{-} &=& E_{1}(n)F_{n-1}^{+}+E_{2}(n-1)F_{n-2}^{+}+  \ldots
+E_{n}(n)F_{0}^{+} \\
F_{n-1}^{+} &=& T_{1}(n-\frac{1}{2})F_{n-1}^{-}
+T_{2}(n-\frac{1}{2})F_{n-2}^{-}+  \ldots
+T_{n}(n-\frac{1}{2})F_{0}^{-} \;\; .
\eea
Hence:
\bea
\Psi_{r,2} & = & W_{1}^{T}(r) F_{r-1}^{+} + W_{2}^{T}(r) F_{r-2}^{+} + \ldots +
W_{r}^{T}(r) F_{0}^{+} \;\; .
\eea
If we introduce the matrices
\bea
E^{(n)} &=& \left( \begin{array}{cccc}
E_{1}(n) & E_{2}(n) & \ldots & E_{n}(n) \\
\bbbzero & E_{1}(n-1) & \ldots & E_{n-1}(n-1) \\
\vdots & \ldots & \ldots & \vdots \\
\bbbzero & \bbbzero & \ldots & E_{1}(1)
\end{array} \right)  \\
T^{(n)} &=& \left( \begin{array}{cccc}
T_{1}(n-\frac{1}{2}) & T_{2}(n-\frac{1}{2}) & \ldots & T_{n}(n-\frac{1}{2}) \\
\bbbzero & T_{1}(n-\frac{3}{2}) & \ldots & T_{n-1}(n-\frac{3}{2}) \\
\vdots & \ldots & \ldots & \vdots \\
\bbbzero & \bbbzero & \ldots & T_{1}(\frac{1}{2})
\end{array} \right) \\
W^{(r)} &=& \left( \begin{array}{cccc}
W_{1}^{T}(r) & W_{2}^{T}(r) & \ldots & W_{r}^{T}(r)
\end{array} \right)
\;\; ,
\eea
then we are able to write the recursion as a product of
these matrices:
\bea
\left( \begin{array}{c}
F_{n}^{-} \\ F_{n-1}^{-} \\ \vdots \\ F_{1}^{-}
\end{array} \right)
= E^{(n)}
\left( \begin{array}{c}
F_{n-1}^{+} \\  F_{n-2}^{+} \\ \vdots \\ F_{0}^{+}
\end{array} \right)
& \; , \; &
\left( \begin{array}{c}
F_{n-1}^{+} \\ F_{n-2}^{+} \\ \vdots \\ F_{0}^{+}
\end{array} \right)
= T^{(n)}
\left( \begin{array}{c}
F_{n-1}^{-} \\  F_{n-2}^{-} \\ \vdots \\ F_{0}^{-}
\end{array} \right)
\eea
\bea
\Psi_{r,2} & = & W^{(r)} \left(
\begin{array} {c} F_{r-1}^{+} \\ F_{r-2}^{+} \\ \vdots \\ F_{0}^{+}
\end{array} \right) \;\;\; .
\eea
Finally, we can write the singular vector in the form\footnote{It is
worth remarking that $P^{(r)} E^{(r)}$ does not contain $E_{k}(r)$.}:
\bea
\Psi_{r,2} & = & W^{(r)} T^{(r)} \left\{ 1+ P^{(r)}
E^{(r)}T^{(r)} + ( P^{(r)}E^{(r)}T^{(r)})^{2} +
\ldots + ( P^{(r)}E^{(r)}T^{(r)})^{r-1} \right\}
\left( \begin{array}{c} 0 \\ 0
\\ \vdots \\ 0 \\ F_{0}^{-} \end{array} \right) \nn
\\ &=& W^{(r)} T^{(r)}\; (1-P^{(r)} E^{(r)} T^{(r)})^{-1}\;
\left( \begin{array}{c} 0 \\ 0
\\ \vdots \\ 0 \\ F_{0}^{-} \end{array} \right) \;\; ,
\eea
where $P^{(r)}$ is the $r$ by $r$ matrix
\bea
P^{(r)} &=&
\left( \begin{array}{ccccc}
\bbbzero & \bbbone & \bbbzero & \ldots & \bbbzero \\
\bbbzero & \bbbzero & \bbbone & \ldots & \bbbzero \\
\vdots & & & & \vdots \\
\bbbzero & \bbbzero & \bbbzero & \ldots & \bbbone \\
\bbbzero & \bbbzero & \bbbzero & \ldots & \bbbzero \end{array} \right) \;\; .
\eea

\section{Conclusions}
We have shown that the ``fusion method'' to derive singular vectors
can be applied successfully for the $N=2$ (untwisted) algebra. We
have given explicit expressions for all the charged singular vectors
$\Psi_{k}$ and for some of the uncharged singular vectors
$\Psi_{r,2}$, and have conjectured the fusion which can be taken to
find every uncharged singular vector $\Psi_{r,s}$. However the
construction is getting rather complicated. Alternatively, one can use
the ``analytic continuation method'' in the spirit of ref.
\icite{adrian1} to find product formulae for $\Psi_{r,s}$ in terms of
$\Psi_{n,2}$ expressions. We were not able to start from a charged
singular vector and use a similar procedure. The ``fermionic''
character of a charged singular vector is responsible for the fact
that the equations (\ref{eq:j_f}) lose terms which are necessary
to obtain the recursion formulae.

\acknowledgements
I am very grateful to my PhD supervisor Adrian Kent for many
discussions about various aspects of this work. I would like to
thank Matthias Gaberdiel and G\'{e}rard Watts for explaining $N=2$
fusion and the $N=1$ Ramond singular vectors to me. I also acknowledge
useful discussions with Zoltan Bajnok about the ``fusion method''.
Furthermore I am very grateful to the referee for pointing out
the possibility of changing the fusion point. Finally, and especially,
my thanks go to Val\'{e}rie Blanchard for proofreading the
preprint.
This work was financially supported in part by SERC and by Greta \& Ludwig
D\"{o}rrzapf.

\end{document}